\documentstyle[11pt,aaspp4]{article}
\tighten
\def\s{^{\rm s} }
\def\min{^{\rm m} }

\def\cc{\,{\rm cm}^{-3} }
\def\hcc{\,h^{1/2}\,{\rm cm}^{-3} }
\def\hmpc{\,h^{-1}\,{\rm Mpc} }
\def\hkpc{\,h^{-1}\,{\rm kpc} }
\def\kms{\,{\rm km\, s^{-1}} }
\def\kmsm{\,{\rm km\, s^{-1}\, Mpc^{-1}} }
\def\keV{\,{\rm keV} }

\def\ergcms{\,{\rm erg\, cm^{-2}\, s^{-1}} }
\def\ergcmshz{\,{\rm erg\, cm^{-2}\, s^{-1}\, Hz^{-1}} }
\def\hergs{\,h^{-2}\,{\rm erg\, s^{-1}} }
\def\msun{\,{\rm M}_\odot }
\def\hmsun{\,h^{-1}\,{\rm M}_\odot }

\def\Kel{\,{\rm K} }
\def\Jy{\,{\rm Jy} }
\def\JyK{\,{\rm Jy/K} }
\def\mJy{\,{\rm mJy} }
\def\pJy{\,{\rm Jy}^{-1} }
\def\mK{\,{\rm mK} }
\def\uK{\,\mu{\rm K} }

\def\GHz{\,{\rm GHz}}
\def\eg{{\it e.g., }}
\def\ie{{\it i.e., }}
\def\etal{{\it et al.}}

\def\avrg#1{{\langle #1 \rangle} }
\def\expo#1{\times 10^{#1}}
\def\arcm{\arcmin\ }
\def\eins{{\it Einstein }}
\def\rosat{{\it ROSAT }}
\def\ginga{{\it Ginga }}
\def\asca{{\it ASCA }}
\def\exosat{{\it EXOSAT }}
\def\heao1{{\it HEAO-1 }}
\def\tenma{{\it TENMA }}
\def\MLT{ $\mbox{M}-[\mbox{L} + \mbox{T}]/2$ }
\def\LTD{ $\mbox{L} - \mbox{T}$ }
\def\LTA{ $(\mbox{L} + \mbox{T})/2$ }
\def\ML{  $\mbox{M}-\mbox{L}$ }
\def\MT{  $\mbox{M}-\mbox{T}$ }

\lefthead{Myers et al.}
\righthead{Sunyaev-Zeldovich Effect in}

\slugcomment{To appear in the {\it Astrophysical Journal}}

\begin{document}

\title{Measurements of the Sunyaev-Zeldovich Effect in the Nearby Clusters 
  \\ A478, A2142 and A2256}

\author{S.T. Myers\altaffilmark{1}, J.E. Baker\altaffilmark{2}, 
A.C.S. Readhead,  E.M. Leitch}
\affil{Owens Valley Radio Observatory, 105-24\\
Caltech, Pasadena CA 91125}
\and
\author{T. Herbig}
\affil{Princeton University, Physics Department, Jadwin Hall \\
P.O. Box 708, Princeton, NJ  08544 }

\altaffiltext{1}{current address Dept. of Physics and Astronomy, 
David Rittenhouse Laboratory, University of Pennsylvania, 209 S. 33rd St., 
Philadelphia, PA 19104-6396}

\altaffiltext{2}{current address Dept. of Astronomy, Campbell Hall, 
University of California, Berkeley, CA  94720}

\begin{abstract}

An X-ray flux limited sample of nearby clusters of galaxies has been defined
for observations of the Sunyaev-Zeldovich effect (SZE) to be carried out on
the Owens Valley 5.5-meter telescope at 32 GHz.  The X-ray sample selection
minimizes the systematic errors introduced by cluster elongation in the
determination of $H_0$.  Due to their proximity, these clusters are
well-studied in the X-ray wave-bands.  The measurement of the SZE in three of
these clusters is reported in this paper: $\Delta T = -375\pm28\uK$ (A478),
$-437\pm25\uK$ (A2142) and $-243\pm29\uK$ (A2256).  These values have been
corrected for radio source contamination, but have not been corrected for the
beam dilution and switching (which are model-dependent).  There is an
additional overall calibration uncertainty of $7\%$.  If the temperature
profile of the clusters is known, then the SZE provides a direct probe of the
baryonic mass in the hot ionized phase of the medium.  We find surface
baryonic mass densities of $(7.5\pm2.5)\times10^{13}\,M_\odot\,{\rm Mpc}^{-2}$
within the $7\farcm35$ FWHM Gaussian beam of the 5.5-m telescope projected on
the cluster centers.  For A2142, A2256, and the Coma cluster previously
observed by Herbig \etal \markcite{herbig95} (1995), we find a consistent
value for the ratio of the SZE determined baryonic mass to the gravitational
binding mass of $M_{sze}/M_{tot}=0.061\pm0.011\,h^{-1}$.  Note that this is a
{\it lower limit} on the total baryon fraction, as there may be significant
contributions from other baryons.  Comparison with the standard big-bang
nucleosynthesis prediction $\Omega_B h^2 = 0.013\pm0.02$ gives a value for the
cosmological density parameter of $\Omega_0 h \lesssim 0.21\pm0.05$, assuming
our limit on the baryon fraction in clusters applies to the Universe as a
whole.  This density is in agreement with independently determined values from
large-scale structure studies.  Recent values for $\Omega_B h^2$ based upon
deuterium abundances are outside the previously accepted range, and in
combination with our data lead to significantly higher or lower $\Omega_0$.
Finally, we present preliminary determinations of the Hubble constant using
X-ray models gleaned from the literature.  The data from the three clusters,
along with the results previously obtained using the same telescope on the
Coma cluster, yield a sample average value $H_0=54\pm14\kmsm$.  We discuss the
uncertainties in these results and future prospects for this method.

\end{abstract}


\section{Introduction} \label{intro}

The scattering of the cosmic microwave background radiation by hot gas in
clusters of galaxies, known as the Sunyaev-Zeldovich Effect (SZE), has been
recognized for over two decades as a potentially important tool for
cosmological and astrophysical studies (Sunyaev \& Zeldovich \markcite{sz80}
1980).  In the SZE, inverse Compton scattering boosts the microwave background
photons to higher frequencies, up-shifting and distorting the Planck blackbody
spectrum.  At low frequencies where the spectrum rises with frequency, this
reduces the intensity of the CMB at a given frequency.  At higher frequencies
where the spectrum falls with frequency, the SZE increases the intensity.
Relative to the background, clusters look dark at low frequencies and bright at
high frequencies.  

A particularly important application of SZE observations is the determination
of the Hubble constant, $H_0$.  Classical determinations of $H_0$ rely upon
the cosmic distance ladder (\eg Tully \markcite{tully88} 1988) and are
therefore subject to the uncertainties inherent in each step of the ladder.
For this reason ``direct'' determinations of $H_0$ would be very important if
the systematic errors could be understood and allowed for.  Examples of such
direct methods are expanding supernova photospheres, gravitational lenses and
the SZE.  For recent determinations of $H_0$ using classical methods, see
Kennicutt, Freedman \& Mould \markcite{kenn95} (1995) and references therein,
as well as Sandage \markcite{sandage96} (1996), and Mould \etal
\markcite{mould95} (1995).

The high temperature ionized cluster medium produces both Compton scattering
(SZE) and thermal brem\-sstrahlung (X-ray) emission that depend upon different
powers of the electron density and temperature, $n_e\,T_e$ for the SZE, and
approximately $n_e^2\,T_e^{1/2}$ for the brem\-sstrahlung component of the
X-ray.  The integral equations for the observed SZE and X-ray
brightness can, given a suitable model for the density profile and knowledge of
the electron gas temperature, be solved for central density and the linear core
radius --- when combined with the observed angular core radius this yields a
value for the angular diameter distance $D_a$.  For clusters at low redshift,
this yields the Hubble constant $H_0$.  For a set of clusters covering a wide
range of redshifts, the determination of $D_a(z)$ as a function of redshift $z$
can constrain or determine other cosmological parameters such as the density
parameter $\Omega_0$ and cosmological constant $\Lambda_0$.

The gas in a massive galaxy cluster has a temperature of roughly $10^8\Kel$ and
a central density in excess of $10^{-3}\cc$.  This leads to an expected
microwave decrement along the line of sight through the cluster center in the
range $0.1\mK$--$1\mK$.  Despite the large size of the SZE relative to
intrinsic anisotropy signals ($\lesssim 50\uK$), observations of the effect
have proven difficult and have been plagued by systematic errors.  However, a
number of experiments have now produced reliable SZE measurements.  Using the
40-m telescope at Owens Valley Radio Observatory (OVRO) at $20\GHz$, Birkinshaw
\markcite{birk90} (1990) found $\Delta T_{RJ}$ of $-444\pm65\uK$, 
$-301\pm49\uK$, and $-354\pm 43\uK$ for the clusters 0016+16, Abell 665, and
Abell 2218, respectively.  Herbig \etal \markcite{herbig95} (1995) measured a
decrement of $-308\pm 51\uK$ in the Coma cluster using the OVRO 5.5m
telescope.  Recent interferometric measurements of the SZE in A2218 (Jones
\etal \markcite{jones93} 1993) and A773 (Grainge
\etal \markcite{grainge94} 1994) using the
Ryle telescope at 15 GHz have been reported.  Interferometric measurements of
0016+16 and A773 have also been made by Carlstrom, Joy \& 
Grego \markcite{carlstrom} (1996) with the OVRO Millimeter Array outfitted 
with 32
GHz receivers.  The SZE in the cluster A2163 has been measured using a
bolometer array on the CSO (Wilbanks \etal \markcite{wilbanks94} 1994) --- 
this measurement of the effect
as an increment on the high-frequency side of the blackbody peak is the first
to place interesting limits on the peculiar motion of massive clusters.

These observations have all been made on fairly distant clusters.  This has
been dictated by the use of large single dishes or large interferometers.
These instruments have small switching angles or small fields-of-view that
resolve out the structure on the angular scales corresponding to nearby
clusters.  In the determination of $H_0$, the models of the X-ray emitting gas
in the cluster potential are just as important as the SZE measurements.
Accurate models are available only for the nearby clusters.  Also, the control
of biases to the $H_0$ determination, such as cluster elongation or
substructure, requires the systematic study of a complete sample of carefully
selected clusters.  At the present, the best samples for this study are drawn
from nearby ($z<0.1$) X-ray selected cluster catalogues.

The newly-completed 5.5-m telescope at OVRO is an ideal instrument for SZE
measurements in a sample of nearby clusters.  The telescope operates at
$32\GHz$ and has a sensitive wide-bandwidth HEMT receiver.  The primary beam
is $7\farcm35$ (FWHM) and the dual-horn switching angle is $22\farcm16$.  This
telescope has the right beamwidth and sensitivity to measure the SZE in
clusters with angular core radii in the range 1\arcm to 22\arcm with
reasonable efficiency.  The Herbig \etal \markcite{herbig95} (1995)
measurements of the Coma cluster ($z=0.023$) recovered approximately 61\% of
the central decrement. 

We report here the first SZE observations undertaken with the specific aim of
estimating and, where possible, eliminating the major known sources of
systematic error in the determination of $H_0$ by means of SZE observations
of a sample of nearby clusters.

\section{The X-ray Cluster Sample} \label{sample}

There are three potential problems in SZE determinations of $H_0$: elongation,
density substructure, and non-isothermality of the cluster gas.  If a cluster
is elongated along the line of sight, then $H_0$ will be underestimated, while
if the cluster is elongated in the plane of the sky, then $H_0$ will be
overestimated.  Clumping of the intracluster medium also causes a bias, due to
the different dependences of the X-ray emission and SZE upon the electron
density and temperature.  In this case, $H_0$ is overestimated by the factor
$\langle n_e^2 \rangle/ \langle n_e \rangle^2$, a quantity which is always
greater than unity.  The measurement of $H_0$ is proportional to $T_e^{3/2}$,
so errors in temperature or temperature gradients will be a problem.  Detailed
discussion of these problems have been presented in the literature
(e.g. Birkinshaw \etal \markcite{birk91} 1991).

The uncertainties due to substructure and isothermality can only be resolved
through high resolution X-ray observations and detailed modeling.  For this to
be possible, nearby clusters must be chosen --- new measurements by \asca and
\rosat of these objects are revolutionizing our understanding of cluster
astrophysics.  However, if clusters are prolate or oblate ellipsoids, axial
ratios of $0.5$ imply an uncertainty of up to a factor of 2 in the
line-of-sight distance through the cluster.  This uncertainty will average out
in determinations of $H_0$ if the SZE is measured in a large enough orientation
unbiased sample of clusters.  A sample chosen by central X-ray surface
brightness will be systematically biased towards clusters that have long axes
along the line of sight.  Such a bias arises because these clusters will have a
greater central brightness than clusters at the same distance with their short
axes along the line of sight.  To guard against this selection effect it is
necessary to choose a complete flux-limited parent sample while staying
sufficiently above the sample flux limit to minimize the biasing effect upon
the selection.

We selected the X-ray flux-limited sample of Edge \etal \markcite{edge90}
(1990) for our parent sample.  This sample is complete for fluxes in the 2--10
keV band $f_x \geq 3.1\times10^{-11}\ergcms$, and is 70\%--90\% complete at
$f_x \geq 1.7\times 10^{-11}\ergcms$.  This sample was derived by Edge \etal\ 
from \heao1 and {\it Ariel V} surveys and cross-checked with \eins and \exosat
observations.

We restrict our observations to the higher flux limit $f_x \geq
3.1\times10^{-11}\ergcms$, galactic latitude $|b| \geq 20\arcdeg$, and $\delta
\geq -23\arcdeg$.  A high-luminosity sub-sample was selected with $L_x \geq
1.25\times10^{44}\hergs$ (2--10 keV), where $H_0 = 100\,h\kmsm$.  The 11
clusters in this subsample are listed in Table~\ref{tbl:sample}.  The core
radii for these clusters fall within the optimum size range for the 5.5-m
telescope. Also, given the X-ray parameters for these clusters, we expect SZE
decrements in the range $250\uK$ -- $750\uK$, which are easily observable with
our system sensitivity.  The redshifts and thus the luminosities listed in
Table~\ref{tbl:sample} were taken from the Edge \etal \markcite{edge90} (1990)
paper.

The Edge \etal\ sample for $f_x \geq 3.1\times10^{-11}\ergcms$ is
complete for $z \leq 0.066$ at $L_x \geq 1.5\times10^{44}\hergs$.  The
lowest luminosity cluster in our sample is Coma ($L_x =
1.85\times10^{44}\hergs$) --- at this luminosity, the sample would be complete
for $z \leq 0.073$.  The most distant cluster in our sample is A2142
($z=0.09$); at this redshift the sample would be complete for $L_x \geq
2.82\times10^{44}\hergs$.  In order to be able to observe a reasonably
large number of clusters, we choose to use the entire sample, though
it is flux-limited only, and not volume-limited out to $z=0.1$.  This
means, therefore, that we are prone at the lowest luminosity levels to 
selection effects such as elongation, and some care will have to be taken
in the interpretation of the statistical results.

From the parent sample, we have selected for our first observations clusters
that are free of strong radio sources which would contaminate our SZE
measurements at $32\GHz$.  The clusters A478 and A2142, massive clusters with
well-measured X-ray profiles, were ideal targets, with large expected SZE
decrements.  The cluster A2256 is optimal because of its high declination.
The Coma cluster was previously observed with this instrument by
Herbig \etal \markcite{herbig95} (1995), and has been included in the
sample.

\section{Observations}\label{obs}

Although SZE signals are roughly an order of magnitude larger than the limits
that have been placed on intrinsic anisotropy signals, the removal of
systematic effects from SZE data still requires great care.  The SZE
measurements must be very sensitive, with uncertainties less than 10\% in
order to contribute less than 20\% to the uncertainty in the estimate of
$H_0$.  Most clusters are tracked over large angular distances across the sky.
The variation of azimuth and zenith angle can introduce significant ground
spillover effects, which can be minimized in intrinsic anisotropy observations
by observing only fields near the North Celestial Pole.  Our observations of
clusters use a three-level differencing technique to remove systematic effects
from ground spillover, the atmosphere, and the receiver.

The observations reported here were made between 15 July 1993 and 6 March
1994.  The 5.5-m telescope at OVRO was used, employing a receiver
operating at a center frequency of 32~GHz, with a primary beam FWHM
of $7\farcm35$.  The receiver noise level for a measurement
was approximately $0.9mK$ in one second of total integration time.
The noise level including atmosphere and ground pick-up is about 50\%
higher than this.  The receiver is discussed in \S~\ref{obs-diff} and
the sensitivity and noise performance in \S~\ref{obs-edit}.

This section contains detailed descriptions of the observing, calibration,
data editing, filtering and analysis procedures.  Much of the discussion in
may be skipped by the casual reader, though it is suggested that the reader
look at the discussion of the referencing in \S~\ref{obs-ref} and the data
analysis in \S~\ref{obs-anal} before proceeding to the presentation of the
results for the cluster observations in \S~\ref{sze-results}.

\subsection{First and Second Differencing} \label{obs-diff}

The receiver has two horns: ANT (antenna) and REF (reference).  These have
nearly identical Gaussian on-sky response patterns with $7\farcm35$ FWHM and a
separation of $22\farcm16$.  The rms pointing accuracy of the 5.5-m telescope
was found to be better than $0\farcm5$ from observations of bright calibration
sources.  The receiver switches between the two beams every
1 ms ($f = 500\,\mbox{Hz}$) by means of a pulse latched switch.
The output of the switch then passes through an amplifier chain before
detection.  Our observations employed as the first stage a TRW InSb HEMT
amplifier with a bandwidth of $5.7$~GHz centered at 32~GHz.  After detection,
the power measurements taken in each switch orientation are accumulated and
the mean difference and the standard deviation of the mean difference are
computed for each $1^{\rm s}$ fundamental integration period.  This first
level of switching removes the offsets and signal common to both beams, and
cancels the offsets due to low-frequency gain fluctuations in the amplifier
chain.

Another level of switching is required to remove gradients.  The telescope is
moved in azimuth by the switching angle $22\farcm16$ and a second difference
is formed.  The period for this switching is $20^{\rm s}$ to $50^{\rm s}$.
The second differencing removes gradients in the atmospheric emission that are
stable on these timescales.

The two 5.5-m beams of the telescope cease to intersect at a height of 853
meters for the angle of $22\farcm16$.  For cloudlets moving through the beams,
a characteristic speed of 1 m/s transports the fluctuations through the beams
on a timescales of $5^{\rm s}$ at this height.  The 500~Hz switching
``freezes'' these fluctuations, which then cancel if they move through both
the ANT and REF beams.  We expect these fluctuations to cause an increased
noise level in the second difference measurements.

The basic measurement is the ``FLUX'' (see Readhead \etal \markcite{ncp89}
1989 or Myers \etal \markcite{ring93} 1993 for a description of this
procedure).  In a FLUX measurement, the telescope first moves (slew time
$\tau_s$) to a position where the REF beam is pointed at the cluster (position
``ON'') and the ANT beam is $22\farcm16$ away in azimuth (position ``R1''),
where it spends time $\tau_i$ integrating.  The single-differenced power
recorded in this position is denoted ``A''.  The telescope is then moved
($\tau_s$ again) so that the ANT beam points at the cluster (``ON'' again) and
the REF beam is displaced by $22\farcm16$ in the opposite direction (position
``R2'') from the initial ANT position.  Two consecutive integration cycles are
performed in the new position, yielding power measurements ``B'' and ``C''.
Finally, the telescope is moved back to the original position for an
additional integration, denoted ``D''.  If a series of FLUX measurements is to
be performed consecutively on the same target, the subsequent ``A''
integrations are started from the same position as the previous ``D''
integration, without a telescope slew.  Therefore, each FLUX beyond the first
in a series takes a total time
\begin{equation} \label{tauf}
\tau_f = 4\,\tau_i + 2\,\tau_s,
\end{equation} 
with the first FLUX in the series taking $\tau_s$ longer for the first
slew.  Typically, $\tau_s = 12^{\rm s}$ and $\tau_i = 20^{\rm s}$,
so a FLUX lasts $\tau_f = 104^{\rm s}$.  The actual slew time $\tau_s$
depends upon the zenith angle of the observation, increasing from $12^{\rm s}$
for zenith angles greater than $30\arcdeg$, to more than $30^{\rm s}$ close to
the zenith.

The individual integrations are normalized and calibrated such that
they measure the antenna temperature differences $\avrg{T_{ANT} - T_{REF}}$.
We therefore construct an individual FLUX measurement
\begin{equation} \label{flux}
\mbox{FLUX} = \frac{1}{4} ( -A + B + C - D ). 
\end{equation} 
The telescope also records a standard deviation (SD) measurement for each FLUX
by measuring the variance of the $1\s$ difference measurements during the
integrations.  This procedure subtracts the average of the signals in two
reference positions on either side of the cluster from the signal on the
cluster.  In terms of antenna temperature on the sky,
\begin{equation} \label{fluxtemp}
\mbox{FLUX} = T_{ON} - \frac{1}{2} ( T_{R1} + T_{R2} ).
\end{equation} 

We can also rearrange the individual integrations A--D, which are recorded 
separately as part of the data log, to form other quantities of interest.
For example, we define the {\it switched difference} 
\begin{equation} \label{SW1}
\mbox{SW1} = \frac{1}{4} ( -A - B + C + D ). 
\end{equation}  
Because the differences $(D-A)$ and $(C-B)$ appear in the SW1, and are the
integrations at the same position, the signal from the sky in the far-field is
cancelled, leaving only time variations of the atmosphere (possibly drifting
through the slightly divergent beams), and spurious signals not common to the
integration pairs.  There is another combination of the integrations that
cancels out the far-field signal,
\begin{equation} \label{SW2}
\mbox{SW2} = \frac{1}{4} ( -A + B - C + D ),
\end{equation}  
which also acts like a time-difference filter, but with twice the frequency
of the SW1.  The more moderate filtering provided by the SW1 has proved to be a
useful diagnostic for periods of bad data, without any bias due to strong
celestial sources in the beams.  Note also that, because of the form of
(\ref{SW1}) and (\ref{SW2}), the SD from the FLUX applies to these quantities
also.  However, unlike the FLUX, the SW1 and SW2 are only single-differences in
time, and tend to have correspondingly larger scatter about the mean values
than the FLUXes.  These quantities are useful for throwing out grossly
discrepant measurements but do not provide a stringent filter for more subtle
problems.

As the telescope tracks the cluster across the sky, the reference beams trace
circular arcs around the cluster.  The reference beams are always separated
from the ON beam purely in azimuth.  The position of the reference beams for a
given FLUX is recorded in the form of the parallactic angle ($\psi$), defined
as the angle between the direction of the North Celestial Pole and the
direction of the zenith.  The parallactic angle is 
\begin{equation} \label{pa}
\tan\psi = \frac{\cos{\lambda}\sin{H}}{\sin{\lambda}\cos{\delta} -
\cos{\lambda}\sin{\delta}\cos{H}},
\end{equation}
where $\lambda$ is the geographic latitude of the telescope ($37\arcdeg\,13
\arcm\,55\farcs7$ in the case of OVRO), $\delta$ is the declination of the 
object, and $H$ is the hour angle.  For objects transiting north of the zenith
($\delta >\lambda$), $\psi =180\arcdeg$ at transit, while $\psi = 0\arcdeg$ at
transit for $\delta <\lambda$.  The position angles of the reference beams
relative to the cluster are $\pm 90\arcdeg$ away from the parallactic angle at
the given time.

The reference beams are (very nearly) symmetric, so that we can fold the
parallactic angle $\psi$ into the range ($-90\arcdeg$, $+90\arcdeg$) by adding
or subtracting $180\arcdeg$ with no loss of information.  We define the
principal parallactic angle ($\psi_p$) in this way.  It is desirable for the
observing time on a cluster to be as evenly distributed as possible over a
large range of principal parallactic angle, since this makes it easier to
identify contaminating sources in the reference beams.  However, clusters
which transit near the zenith spend the majority of their time at the extremes
of their $\psi_p$ range.  The $7\farcm35$ beam FWHM at the switching
separation of $22\farcm16$ subtends $\Delta\psi_p \sim 19^\circ$; this is the
effective `resolution' in parallactic angle.

\subsection{Lead-Trail Referencing} \label{obs-ref}

The first and second levels of switching produce a FLUX measurement which is a
``double-difference'': the difference between the signal in a central
$7\farcm35$ FWHM beam and the average of two reference beams $22\farcm16$ on
either side of the central beam.  However, even with this differencing,
offsets in the FLUX levels remain at the $\sim 100\uK$ level.  These offsets
change as the telescope tracks a cluster, therefore we must impose another
level of differencing to remove this systematic effect.  

The third level of switching involves observations of LEAD and TRAIL control
fields far from the cluster centers.  The LEAD and TRAIL fields are offset by
$\pm\delta\alpha$ in right ascension from the cluster.  Observations of the
LEAD, MAIN, and TRAIL fields are separated by $\delta\alpha\min$ in time so
that the telescope tracks through the same azimuth and zenith angles for each
field.  This referencing was used in the Coma observations of Herbig \etal
\markcite{herbig95} (1995).
From each MAIN FLUX, we subtract the average of the LEAD and TRAIL FLUXes to
form the referenced field \MLT.  This level of switching corrects for offsets
dependent upon telescope orientation such as curvature in the ground
spillover.  We generally choose $10\min\leq\delta\alpha\leq 25\min$, which is
large enough so that the efficiency of the observations is not greatly reduced
by frequent slewing, but small enough that temporal variations in the offsets
are not severe.  Though essential for the removal of systematic errors, the
three-level switching technique employed greatly reduces the efficiency of our
observations.  Only one-sixth of the observing time is actually spent on the
clusters, while the remainder is spent on the reference arcs and control
fields.

The criterion for selecting $\delta\alpha$ is that both LEAD and TRAIL fields
should be free of confusing radio sources bright enough to adversely affect
the observations.  We avoided sources found in the $4.85\GHz$ Green Bank
survey (Gregory \& Condon \markcite{gregcond91} 1991).  For clusters with
$\delta\alpha =15\min$ a typical scan consisted of six FLUXes on each of the
LEAD, MAIN, and TRAIL fields.  Each FLUX had a total integration time $4\tau_i
= 92\s$.  Fewer FLUXes could be completed per $15\min$ scan near the zenith
because of the increased slew time there.  The cluster A478 was observed with
$\delta\alpha =19\min$, which allowed ten FLUXes per scan with integration
time $4\tau_i =68\s$.  Approximately three minutes per scan were allocated for
calibration and slewing.  To ensure complete parallactic angle coverage for
all three fields, observations were scheduled in different LST blocks.  These
different schedules filled in gaps that would have been created if the
telescope had been slewing at the same LST in every schedule.  The pointing
positions (J2000) used for the cluster LEAD, MAIN, and TRAIL fields in the
observations are listed in Table~\ref{tbl:pos}.

\subsection{Calibration} \label{obs-cal}

Calibration of the telescope and receiver was conducted regularly during the
period from 1993 to 1995, with the most exhaustive calibration measurements
taken in 1995.  We derive our calibration scale from the 1995 data and
apply it to our SZE data taken in 1993 -- 1994 using observations of
standard sources.

The telescope records the differenced power measurements for each FLUX in
units of counts.  In addition, the power from a calibrated noise diode is
measured once at the end of each scan to normalize the changing gains in the
system.  The noise diode is calibrated in antenna temperature units by
comparing its power output with hot (300 K) and cold (77 K) loads filling the
beams, and in flux density units (1 Jansky = $10^{-23}\ergcmshz$) by
comparing against `standard' radio sources such as Jupiter, Mars,
and DR21.

In addition to the waveguide switch that switches between the ANT and REF
horns, the 5.5-m receiver has an additional switch in each of the ANT and REF
arms that allows us to switch between the horn and a cold load as inputs (SKY
and LOAD).  The value of the CAL is measured with both arms looking at the
loads, since this improves the stability of the measurement.  

For each FLUX measurement, we assign a CAL value by linearly interpolating
between the CAL measurements made before and after each FLUX.  However, we do
not use a CAL value if its standard deviation is greater than 5\% of its value.
This criterion rejects only about 1\% of the CALs.  We also do not use data if
the difference between the bracketing CALs is greater than 5\% of the mean
value.  Each FLUX and SD (standard deviation) measurement is divided by its
assigned CAL value and multiplied by the appropriate temperature equivalent 
$T_{cal}$ to yield antenna temperature.

The receiver system is slightly nonlinear.  This nonlinearity has been
measured and shown to be stable.  Herbig \markcite{herbig93} (1993) has
modeled the nonlinearity by assuming that the observed output power difference
and the true input power difference are related by a linear function of the
observed power.  Because the CAL measurements are made with the load switches
in the LOAD position, while the FLUX measurements are made with the switches
in the SKY position, a correction must be made for the nonlinearity difference
between the two power levels.  We use the average total power, which is
recorded once during each scan with the antenna looking first at the sky and
then at the load, to correct the CAL value to the power level of the FLUX
measurement.  Since the power levels change somewhat over the course of the
observations, the correction must be applied to each FLUX individually.  For
the dataset as a whole, over a range of zenith angles and atmospheric noise
levels, the correction factor ranged from $0.96$ to $1.00$.

The aperture and beam efficiencies of the antenna are determined by comparing
the flux densities of `standard' sources to the measured antenna temperatures.
Measurements of Jupiter, Mars, DR21, NGC7027 and 3C286 were made with the
5.5-m 32 GHz system during the period from February to June 1995.  The flux
density scale adopted for the 32~GHz observations are based upon a physical
temperature of $T_J = 144 \pm 8$~K (Wrixson \etal \markcite{wrixson71} 1971)
for Jupiter at this frequency.  This value encompasses the various measured
temperatures for Jupiter, which range from $137\Kel$ to $153\Kel$.  DR21 and
Mars are used as secondary calibrators.  DR21 is an H II region in the
galactic plane, and the 5.5-m measurements are confused by emission in the
reference beams.  The measured flux of DR21, relative to Jupiter assuming $T_J
= 144\Kel$, is $S_{DR21} = 19.4\pm0.3 \Jy$, in agreement with $19.4$~Jy
measured by Aller (private communication), though not with the $18.2$~Jy value
given by the Baars scale (Baars \etal \markcite{baars77} 1977).  We also
measure a temperature of Mars relative to Jupiter giving $T_M = 180\pm3 \Kel$.
The flux densities of the planetary nebula NGC7027 and of the radio galaxy
3C286 were also measured relative to DR21.  We find $S_{3C286} = 1.92\pm0.06
\Jy$ for $S_{DR21} = 19.4\Jy$.  Note that NGC7027 is expanding, and during Feb
- June 1995 $S_{N7027} = 5.10\pm0.14\Jy$.  The scale is listed in
Table~\ref{tbl:cals}, both relative to Jupiter and using the adopted absolute
value $T_J = 144$~K.

The equivalent flux density and thermodynamic load temperature for the
CAL diode are related by 
\begin{equation} \label{apeff}
 \frac{S_{cal}}{T_{cal}} = \frac{2\,k}{A_p\,\eta_A} = \frac{116.8\JyK}{\eta_A}
\end{equation} 
where the physical area $A_p$ of the 5.5-m telescope is $23.64\,{\rm m}^2$.
Using the $T_J = 144$~K flux density scale and measurements of the standard
sources and hot and cold loads in April 1993, June
1993, October 1994, and April 1995, we calculate the aperture efficiency of
the 5.5-m telescope to be $\eta_A = 0.497\pm0.007$ ($4.255\pm0.014\mK\pJy$).
The CAL temperature $T_{cal}$ corresponds to the equivalent Rayleigh-Jeans
temperature increase in a blackbody filling the entire telescope beam.
Because the power output of the CAL diode has been found on occasion to vary
on timescales of months, the equivalent temperature or flux density of the CAL
must be determined for the time period of calibration.  During the SZE
observing sessions from July 1993 to April 1994 the ratio of DR21 to the CAL
was measured to be $0.1542\pm0.0022$.  The stated uncertainty is the standard
deviation about the mean for 12 observations spread over this interval and
should reflect the range of possible variation in the CAL over this period.
From these measurements we deduce that the CAL was stable to at least $1.4\%$
rms.  Assuming a flux density of $19.4\Jy$ for DR21, we adopt a constant value
of $S_{cal} = 125.8 \pm 1.8\Jy$, corresponding to $T_{cal} = 535 \pm 11 \mK$.
After normalizing our FLUX values by the CAL, we multiply by $T_{cal}$ in
order to compare the measurements with the expected thermal noise.

The relationship between the flux density and measured antenna temperature
of a source filling the main beam of the telescope is 
\begin{equation} \label{jyperk}
 \frac{S_\nu}{T_a} = \frac{2\,k}{\lambda^2}\,\Omega_{mb} = 161\pm4\JyK
\end{equation} 
for the the measured main beam solid angle $\Omega_{mb} = (5.12 \pm 0.14)
\times 10^{-6}\,\mbox{sr}$ and observing wavelength $\lambda = 0.938\,
\mbox{cm}$.  Note that (\ref{apeff}) and (\ref{jyperk}) differ by 
\begin{equation} \label{bmeff}
 \frac{\eta_B}{\eta_A} = \frac{A_p\,\Omega_{mb}}{\lambda^2} = 1.38 \pm 0.04.
\end{equation} 
where $\eta_B$ is the fraction of the entire telescope beam in the forward
main beam.  From the aperture efficiency and main beam solid angle we derive a
beam efficiency $\eta_B = 0.684\pm0.021$.  A radio source of flux density
$S_{cal}$ is equivalent to a source filling the main beam with uniform
Rayleigh-Jeans brightness temperature
\begin{equation} \label{tbrcal}
   T^\ast_{cal} = \frac{\lambda^2}{2\,k\,\Omega_{mb}}\,S_{cal} = 
   \frac{T_{cal}}{\eta_B} = 782 \pm 24\mK.
\end{equation} 
This is the appropriate CAL value to use for calibrating the astronomical
signal in the main beam.  The FLUX measurements, already scaled by $T_{cal}$,
must be divided by the beam efficiency $\eta_B$ to effectively scale by 
$T^\ast_{cal}$.  

The FLUX and SD values must also be scaled by a factor which accounts for
atmospheric attenuation.  Atmospheric
attenuation reduces the intensity by a factor $\approx \mbox{e}^{-a \sec
\theta_Z}$, where $\theta_Z$ is the zenith angle and $a$ is the optical 
depth.  In good weather at $\nu = 32$ GHz, $a \sim 0.04$ (assuming $a
\sim T_{atm}/T_{phys} \approx 11\Kel / 270\Kel$).  The estimated rms variation
in the atmospheric optical depth is $\sigma_a \lesssim 0.02$, based upon 
water-vapor radiometry data obtained at OVRO.

In summary, the raw FLUX and SD values are first divided by the CAL value,
then scaled by the CAL temperature $T_{cal}$.  After editing, the
measurements are scaled by the correction factor :
\begin{equation} \label{kappa}
\kappa = \frac{e^{a \sec \theta_Z}}{\eta_B} 
       = 1.46\,e^{0.04 \sec \theta_Z}.
\end{equation} 
At this stage, the measurements are in degrees Kelvin, and are equivalent to
the differences in temperature between two high-temperature uniform blackbody
emitters (calculated using the Rayleigh-Jeans formula) filling the main beams
of the ANT and REF horns that would produce the observed power differences.

The total calibration uncertainty is the quadrature sum of the uncertainties
in: (1) the temperature of Jupiter $T_J$ (5.6\%), (2) the ratio of our
measured flux of DR21 relative to Jupiter (1.6\%), (3) the ratio of the CAL
$S_{cal}$ to DR21 during the course of the observations (1.4\%), (4) the main
beam solid angle converting flux density to brightness temperature (2.7\%),
and (5) the atmospheric attenuation variation (2.0\%), giving a total
calibration error budget of $6.9\%$.  The dominant uncertainty is the absolute
brightness temperature of Jupiter.  The relative flux density scale, (2) and
(3), is accurate to $2.1\%$, while the temperature scale, (2), (3) and (4), is
accurate to $3.4\%$ (without atmospheric correction).  For improvement in the
SZE measurements beyond what is reported here, a more accurate absolute flux
density scale at 32~GHz is required.

\subsection{Data Editing} \label{obs-edit}

Much of the data taken during the afternoon or when the weather was bad was
clearly unusable, and a method of editing the data without introducing any
systematic biases had to be devised.  The two general editing methods used
were editing based on the SD values and filtering based on the standard
deviation of the switched difference measurements.

Before applying these methods, we removed FLUXes for which the elapsed time
exceeded the expected duration by more than 4 seconds.  Excessive durations
could be caused by extremely high winds or tracking problems which would
corrupt the data.  In addition, on several occasions problems arose with the
zenith angle encoder which caused the drive to fail, and resulted in a few
extremely long FLUXes during which the telescope was not tracking the source.

The procedure for thermal editing is based upon the expected thermal
noise in a FLUX measurement.  The thermal standard deviation is given by:
\begin{equation} 
\sigma_{th} = \frac{2T_{sys}}{\sqrt{4\tau_i \, \Delta\nu}}
            = \frac{T_{sys}}{\sqrt{\tau_i \, \Delta\nu}},
\end{equation} 
where $4\tau_i$ is the total integration time of the FLUX, 
$\Delta\nu = 5.7$ GHz is the bandwidth of TRW amplifier, and the numerical
factor arises because of the double switching inherent in the FLUX measurement
and the FLUX definition of (\ref{flux}).  The system temperature $T_{sys}$ is
given by
\begin{equation}
T_{sys} = T_0 + T_{atm} \sec \theta_Z. 
\end{equation} 
$T_{atm}$ is the atmospheric emission temperature per airmass.  Included in
$T_0$ are contributions from the receiver, the ground, and the $2.726\Kel$
cosmic background (Mather \etal \markcite{mather94} 1994),
\begin{eqnarray}
T_0 & = & T_{rx} + T_{gnd} + 2.726\Kel \\
T_{rx}  & = & 33.0\pm 1.6\Kel \nonumber \\
T_{gnd} & \approx & 8\Kel \nonumber \\
T_{atm} & \sim & 10\Kel.  \nonumber 
\end{eqnarray}
At one airmass for $\tau_i = 20^{\rm s}$, this gives $\sigma_{th} = 0.16\mK$.
Even with the overhead from the three-level switching, we would expect to
integrate down to a noise level of around $36\uK$ in the referenced field \MLT
in about 3 hours of total integration time.  In practice, additional
atmospheric noise not removed by the double differencing increases the noise
significantly above this level.

We first wish to reject blocks of data which are clearly contaminated by bad
weather.  Our method is a modification of the method developed by Brandt
\markcite{brandt92} (1992), in which one searches for blocks of good data 
which contain a specified number of points, typically 10 to 25, within a
limited time range, typically 2 hours.  Sliding buffers are moved over the
dataset, and any point that is contained within a ``good'' block of data is
accepted.

For each point $i$ in the dataset, we construct the ``test'' statistic
\begin{equation} \label{test}
t_i = \mbox{X}_i / \sigma_{th,i}.
\end{equation}
We have divided the value by the corresponding thermal noise level
$\sigma_{th,i}$ to account for the increased scatter expected at high zenith
angle, where the telescope is looking through a longer column of air, and
to allow comparison of data points with differing integration times.  The
value $X$ used as the statistic can be FLUX, SD, SW1 or SW2.

The test values are placed into ``buffers'' of $N$ consecutive points, for
which the mean and the standard deviation about the mean are calculated.  For
buffer $j$,
\begin{mathletters}
\begin{eqnarray}
\overline{t}_j & = & \frac{1}{N} \sum_{i=j}^{j+N-1} t_i \\
s_j & = & \left[ \frac{1}{N} \sum_{i=j}^{j+N-1} ( t_i - \overline{t}_j )^2
  \right]^{1/2}.
\end{eqnarray}
\end{mathletters}
Each test buffer contains $N$ points, restricted to be taken over a maximum
span of not more than 2 hours.  

The filter is effected by examining all buffers that contain a given data
point $i$.  A data point $i$ is rejected if there exists no buffer $j$ of the
chosen length $N$ containing the given point for which the standard deviation
$s_j$ is less than a chosen cutoff value $s_{max}$ and for which the mean
$\overline{t}_j$ is less than a limit $\overline{t}_{max}$.  We used either
the mean or the standard deviation of the test values $X$ in the
filter, but not both at once.  We designate these filters as ``meanX$[N,
\overline{t}_{max}]$'' or ``sigX$[N,s_{max}]$''.  In addition, we
can simply throw out discrepant values with $t_i > t_{max}$, with filter
designation ``X$[t_{max}]$''.  This brute-force rejection is only useful for
the SD, where it can remove single points with large error bars that escape
previous filters.  In these filters, the FLUX values themselves are not used
to avoid bias of the results.

Through experimentation upon the data, we have developed a filtering 
sequence that produces good results and is robust to slight changes in the
filter parameters:

\begin{enumerate}

\item Filter on mean of SD, with $N=25$ and $\overline{t}_{max}=2.0$ 
(meanSD$[25,2.0]$).

\item Reject points with high SD, using $t_{max}=2.5$ (SD$[2.5]$).

\item Filter on standard deviation of SW1, with $N=10$ and $s_{max}=7.5$ 
(sigSW1$[10,7.5]$).

\end{enumerate}

After our filtering, we are left with a ``clean'' distribution of
FLUXes, although a few FLUXes with discrepant values from the mean may
remain.  These points can escape our culling procedure if they are due to
stationary structures in the far-field atmosphere, or, of course, spurious
noise or interference that happens to mimic the switching scheme of the FLUX
(unlikely, but possible).  The removal of ``bad'' FLUXes which show no ill
effects in the respective SW1, SW2, or SD without biasing the data is a tricky
business.  A conservative approach would be to reject all FLUXes that are part
of a scan on a given field for which the standard deviation of the FLUXes about
the scan mean is above some limit --- this should be equivalent to a
noise-level edit.  In practice, we will use the scan standard deviations to
down-weight this data during the analysis (see next section).

Our approach, one we have adopted with success in the past (Readhead \etal
\markcite{ncp89} 1989, Myers \etal \markcite{ring93} 1993), is to throw out 
the FLUXes on the tail of the FLUX distribution
for a given field using an iterative procedure to ensure that the resulting
post-edit distribution is not skewed by applying too stringent a cutoff.  This
rejection procedure we designate as ``rejX$[\sigma]$''.  For the cluster data,
we reject iteratively the {\it referenced} temperatures $\Delta T_{MLT}$ (see
below) with a cutoff of $4\,\sigma$ (rejMLT$[4]$).  Note that for reasonable
cutoffs ($> 3\,\sigma$), this procedure will not introduce a bias into the
data, since outliers are rejected one at a time and after each iteration the
mean and standard deviation are re-calculated.

Discussion of the effects of different editing and filtering parameters
will be presented in the next section.

\subsection{Analysis of SZE Measurements} \label{obs-anal}

Software was designed to perform the subtraction of the reference LEAD and
TRAIL fields from the MAIN field.  The program first multiplies the filtered
and edited data by the correction constant $\kappa$ from equation 
(\ref{kappa})
\begin{equation} \label{defdt}
\Delta T_i = \kappa\,\mbox{FLUX}_i \qquad\qquad 
  \sigma_i = \kappa\,\mbox{SD}_i
\end{equation}
to convert the power differences into Rayleigh-Jeans temperature differences
in degrees Kelvin between the ANT and REF main beams.  The separate FLUXes from
the adjacent MAIN (M), LEAD (L), and TRAIL (T) scans are then matched based on
their proximity in the azimuth and zenith coordinates.  The FLUXes in the MAIN
fields are matched with the closest FLUXes in the adjacent LEAD and TRAIL
fields that have not already been matched to other MAIN points.  Matching LEAD
and TRAIL measurements are required for each MAIN --- if no reference FLUXes
are found in the LEAD or TRAIL that are closer than 5\arcm (a substantial
fraction of the beamwidth) from the MAIN field, the MAIN FLUX is discarded.

For each triplet, we calculate the ``referenced'' differences
\begin{mathletters}
\begin{eqnarray} \label{mlt}
\Delta T_{MLT} & = & \Delta T_M - \frac{1}{2}(\Delta T_L + \Delta T_T) \\ 
\Delta T_{M-L} & = & \Delta T_M - \Delta T_L \\ 
\Delta T_{M-T} & = & \Delta T_M - \Delta T_T \\ 
\Delta T_{\overline{LT}} & = & \frac{1}{2}( \Delta T_L + \Delta T_T ) \\ 
\Delta T_{L-T} & = & \Delta T_L - \Delta T_T,
\end{eqnarray}
\end{mathletters}
Gaussian error propagation is used to compute the statistical
uncertainties in these differences.  For example, 
\begin{equation} 
\sigma^2_{MLT} = \sigma_M^2 + \frac{1}{4}(\sigma_L^2 + \sigma_T^2). 
\end{equation}

The referenced points are then weighted and the statistics are formed. For each
of the N points with \MLT, \ML, \MT, etc., we form $\overline{\Delta T} \pm
\epsilon$ where
\begin{mathletters}
\begin{eqnarray} \label{stats}
\overline{\Delta T} & = & { 1 \over W_1 } \sum_{j=1}^N { w_j \, \Delta T_j } \\
\epsilon^2 & = & { W_2\over W_1^2} \sigma^2 \\
\sigma^2 & = & { N\over N-1 }\, { 1\over W_2} \sum_{j=1}^N
   w_j^2\, (\Delta T_j-\overline{ \Delta T })^2 \\
W_1 & = & \sum_{j=1}^N w_j \\
W_2 & = & \sum_{j=1}^N w_j^2.
\end{eqnarray}
\end{mathletters}

The weights are formed from a combination of the individual errors $\sigma_j$
(e.g. equation~\ref{defdt}) and the standard deviation of the points within
scans $\sigma_{sc}$,
\begin{equation} \label{weights}
w_j^{-1} = a_0 + a_1\, \sigma_j^2 + a_2\, \sigma_{sc,j}^2.
\end{equation}
The standard deviation SD recorded with each FLUX generally underestimates the
actual error by a factor of 2 or 3.  This is a result of atmospheric
fluctuations on a timescale longer than the duration of the FLUXes.  A better
estimate of the true error is the scatter of the FLUX measurements within a
scan.  An appropriate weighting is $a_1 = a_2 = 1$ and $a_0 = 0$.  The
inclusion of the individual SDs $a_1\neq0$ guards against anomalously low scan
standard deviations causing very large weights, while the weight will normally
be dominated by the larger scan standard deviations $\sigma_{sc,j}$.

Fig~\ref{fig:edit} shows our results for A478, A2142 and A2256 using twenty
different filtering methods.  The referenced measurements $\Delta T_{MLT}$ are
plotted.  Both unweighted (u: $a_0=1$, $a_1 = a_2 = 0$) and weighted (w: $a_1
= a_2 = 1$, $a_0 = 0$) results are shown.  The methods are enumerated by
filtering and outlier editing schemes.  There were five filtering methods
chosen:
\begin{quote}
\begin{description}
\item[Method \phantom{1}1:] No filtering
\item[Method \phantom{1}5:] SD$[3.0]$
\item[Method \phantom{1}9:] meanSD$[25,2.5]$, SD$[3.0]$
\item[Method 13:] meanSD$[25,2.0]$, SD$[2.5]$, sigSW1$[10,7.5]$
\item[Method 17:] meanSD$[25,1.5]$, SD$[2.0]$, sigSW1$[10,6.0]$
\end{description}
\end{quote}
In addition, for each filter, four different outlier rejection schemes
were tested:
\begin{quote}
\begin{description}
\item[Method n+0:] No outlier editing
\item[Method n+1:] rejMLT$[5]$ ($5\sigma$ rejection)
\item[Method n+2:] rejMLT$[4]$ ($4\sigma$ rejection)
\item[Method n+3:] rejMLT$[3]$ ($3\sigma$ rejection)
\end{description}
\end{quote}
These filter methods 1--20 are roughly in increasing order of fraction of the
data rejected.  A total of $802.5$ hours of MAIN, LEAD and TRAIL data were
passed with no filtering, editing, or weighting (Method 1u), while only
$351.7$ hours of data were accepted for Method 20u.

For the weighted means, the form of (\ref{stats}) suggests an effective number
of points
\begin{equation} \label{effdf}
N_{eff} = \frac{W_1^2}{W_2} = \frac{(\sum_{j=1}^N w_j)^2}{\sum_{j=1}^N w_j^2}
\end{equation}
The effective fraction ($\sim N_{eff}/N$) of accepted data ranges from 35\%
(Method 1w) to 27\% (Method 20w) with little variation between outlier editing
modes for each filter method.  Note the smaller variation between the weighted
methods; the weighting largely takes care of the editing by down-weighting bad
stretches of data.  The unedited and unweighted data point (Method 1u) is not
visible in Fig~\ref{fig:edit}, as it is heavily corrupted by bad data and is
off-scale.  The corresponding weighted point (Method 1w) is consistent with 
the other data points.

We adopt method 15w: meanSD$[25,2.0]$, SD$[2.5]$, sigSW1$[10,7.5]$
filtering, rejMLT$[4]$ outlier editing, and weighting $a_1 = a_2 = 1$, $a_0 =
0$ (marked in Fig~\ref{fig:edit}).  Because none of the results differ
significantly with respect to the statistical error bars, we are
confident that even fairly large differences in the filtering method have
negligible effect upon our results.  We conclude that our automatic editing
procedure is reliable and robust.

\section{SZE Results} \label{sze-results}

In all three clusters, A478, A2142, and A2256, we find significant detections
of a microwave decrement that we attribute to the SZE.  Table~\ref{tbl:sze}
lists the measured values of $\Delta T$ for these clusters.  In addition to the
fully referenced difference \MLT, and the single-referenced differences \ML and
\MT, we also list the averages for each of the MAIN, LEAD, and TRAIL fields
separately, as well as the weighted difference and average of the LEAD and
TRAIL.  If there is no significant source contamination, we expect the averaged
\LTD\ to be consistent with zero, provided that our switching technique is
subtracting all significant ground spillover effects.  However, if ground
spillover or atmospheric emission varies with a timescale shorter than the time
between the MAIN, LEAD, and TRAIL scans, the \LTD\ may be inconsistent with
zero, but this effect could be expected to average out over the course of an
observing season.  Source contamination in the control fields will generally
result in non-zero values for \LTD and will contaminate the average \LTA.

The measurements of the SZE as a function of principal parallactic angle
$\psi_p$ are presented in Fig~\ref{fig:a478} -- Fig~\ref{fig:a2256}.  The
upper panels of each plot show the individual MAIN, LEAD and TRAIL binned
averages, while the lower panels show the referenced \MLT\ and \LTD.  The
scatter of these referenced values with parallactic angle can give an
estimate of possible reference arc contamination.

For A478 we find $\Delta T=-375\pm24\uK$ with 64 total hours of integration
time, a $15.6\sigma$ detection.  Fig~\ref{fig:a478} shows $\Delta T$ binned by
parallactic angle $\psi_p$.  The close agreement of the LEAD and TRAIL over the
entire range of $\psi_p$ suggests that there is no significant source
contamination of the control fields.  Although the \MLT\ is nearly constant
with $\psi_p$, with $\chi^2 = 1.70$ about the mean in the 5 bins, the
individual fields in the upper figure show variations of $\sim 400\uK$
resulting from ground spillover.  This demonstrates the necessity of our LEAD
and TRAIL referencing.  The LEAD and TRAIL difference \LTD\ is consistent with
zero, and has $\chi^2 = 6.87$ about zero in the 5 parallactic angle bins, which
is significant only at the 79\% level.  Thus, we conclude that there is no
evidence for source contamination in the A478 data.

A microwave decrement of $\Delta T=-420\pm19\uK$ is found in A2142, a detection
significant at the $22\sigma$ level.  There were 85 hours of usable integration
time on this cluster.  As shown in Fig~\ref{fig:a2142}, the LEAD and TRAIL
track each other with an average offset of $188\uK$.  There is a mean \LTD\
difference of $-66\pm20$, which is marginally significant at the $3\sigma$
level.  In Fig~\ref{fig:a2142} we see a feature in the \LTD\ at parallactic
angle $\psi_p\sim-20^\circ$.  This is consistent with a source in the LEAD
reference arc centered at $\psi_p=-21^\circ$ (see next section).  The 
$\chi^2$ is $3.14$ about zero in the 4 parallactic angle bins, although 
since most of the data is at the ends of the arcs, the effective number
of degrees of freedom (\ref{effdf}) is only $2.7$.  The significance of this
chi-squared value is 67\%.  The referenced \MLT\ has
a $\chi^2$ of $2.1$ about the mean for $1.5$ effective degrees of freedom,
which is significant only at the 75\% level.  

We measure $\Delta T=-218\pm14\uK$ for A2256, a detection significant at
$15.6\sigma$.  There was 310 hours of integration time on A2256.  The MAIN,
LEAD, and TRAIL data are shown binned by parallactic angle in
Fig~\ref{fig:a2256}.  Because A2256 is circumpolar, the magnitude of the
average LEAD and TRAIL is somewhat smaller for this cluster as it traverses a
more restricted range in azimuth and zenith angle and is not as strongly
affected by changes in ground spillover as are A2142 and A478.  The overall
average difference between LEAD and TRAIL is again consistent with zero,
$-22\pm15\uK$.  However, when the data is binned in parallactic angle as in
Fig~\ref{fig:a2256}, we find $\chi^2=7.88$ versus zero for $4.3$ effective
degrees of freedom, which is significant at the 88\% level.  After
referencing, the $\chi^2$ of \MLT\ data versus the mean is $5.17$ for
$3.3$ effective degrees of freedom, or 81\% significance.

In all three cases, the MAIN field is clearly showing the SZE decrement
relative to the LEAD and TRAIL, while all three track together with
parallactic angle.  The variation of the LEAD and TRAIL average offset 
with $\psi_p$ shows the importance of the extra referencing to remove the
differential ground spillover component.  This referencing was also performed
using a slightly different procedure in the Coma observations of Herbig \etal
\markcite{herbig95} (1995).

The dataset is also robust with respect to separating into different time
periods and different times of day.  No trends are seen that are significant
compared to the statistical uncertainties.

After referencing, we have rms uncertainty levels of $14\uK$--$24\uK$.  On
these angular scales at this fluctuation level, anisotropies in the cosmic
microwave background radiation itself can be expected to be detectable.
Intrinsic fluctuations on this scale are expected to be in the range
$5\expo{-6} \lesssim \Delta T/T \lesssim 2\expo{-5}$ ($14\uK$--$55\uK$) in
the most popular models (eg. Bond \etal \markcite{bond94} 1994).  Our
instrumental filtering can be expected to reduce the CMB anisotropy signal
somewhat, as will the smearing with parallactic angle of the reference beams.
The clearest indicator of CMB fluctuations would be significant \LTD\
differences; the fact that we have no clear detection of a LEAD -- TRAIL
difference, except possibly in A2142, indicates that the true anisotropies are
not much greater than the predicted range.  In a separate observing program,
we have conducted a microwave background anisotropy experiment using the 5.5-m
telescope and the same instrumental configuration, the results of which will
be reported in an upcoming paper.  If the background fluctuations are indeed
in the expected range, then pushing the SZE on these scales to much lower
noise levels will not be possible using single-frequency measurements such as
ours. Note, however, the SZE in more distant and hence much smaller
angular-sized clusters will not be so badly affected, as the background
fluctuations on smaller scales are expected to be considerably smaller.

A significant contribution to systematic error in the SZE measurements is
foreground contamination by galactic and extragalactic emission.  At this
frequency and angular scale we believe that galactic dust and free-free
emission are not likely to be major contaminants.  However, synchrotron
emission by discrete extragalactic sources is known to be a significant
problem.

\section{Source Contamination} \label{src}

An unfortunate aspect of centimeter wavelength observations of the SZE is that
they must contend with the presence of radio sources that can mimic or hide the
effect.  High resolution radio maps must be made at or near the frequency of
observation to deal effectively with this problem.  Unfortunately, many radio
sources are also variable, and if a cluster is contaminated by such sources, be
they field object or associated with the cluster, simultaneous observations
must be made on different telescopes.  This was not feasible for the present
work; however, previously published observations of the fields of many of our
clusters suggest that our results are not greatly affected by source
contamination.  Note that in this respect interferometric SZE observations
are superior --- the longer baselines provide the simultaneous high-resolution
information necessary for source identification, and if the baselines extend
far enough compared to the shortest spacings, a clean subtraction of the
interfering sources can be made.  However, with single-dish data, we have
not this luxury.

Radio observations of adequate resolution for computing corrections to SZE
measurements have been published for many of the clusters in our sample.
However, these maps extend only about $20\arcm$ from the cluster centers, so
the possibility exists that there are unrecognized contaminating sources in
the reference arcs.  For the LEAD and TRAIL fields, we use data from the 1987
Green Bank survey at $4.85$ GHz (Gregory \& Condon \markcite{gregcond91}
1991), which lists sources down to a flux limit of 25 mJy.  Unfortunately, the
declination range of this survey does not include A2256, and the radio
environments of the LEAD and TRAIL field for this cluster are at present
unknown.  However, the excellent match between the LEAD and TRAIL fluxes and
the stability of the referenced $\Delta T_{MLT}$ with $\psi_p$ give us
confidence that contamination is not a serious problem in A2256.  Many of the
clusters in our sample fall below the southern declination limit of the Green
Bank survey, and for these we obtained source positions from preliminary
results of the southern Parkes-MIT-NRAO (PMN) survey (Griffith \& Wright
1994).

\subsection{40-m Observations at OVRO} \label{src-ovro}

The published observations of discrete radio sources have been carried out at
frequencies considerably lower than our observing frequency of $32\GHz$.  
For this reason, we observed the contaminating sources from
the Green Bank and PMN surveys with the OVRO 40-m telescope at $18.5\GHz$.  We
observed sources within $30\min$ in right ascension and $32\arcm$ in
declination from the cluster center, in order to cover the possible
locations for the LEAD and TRAIL fields.  Sources near the cluster centers were
also identified from higher resolution radio maps.  Images of A478 and A2142
at $2.$7, $4.75$, and $10.7\GHz$ were available (Andernach \etal 
\markcite{ander86} 1986), as well as for A2256 at 610 and 1415 MHz 
(Bridle \etal \markcite{bridle79} 1979).

At a frequency of $18.5\GHz$, the 40-m telescope has a beamwidth of $2\arcm$
FWHM.  Because of its larger collecting area and much higher sensitivity to
point sources, the 40-m can measure any sources that would be bright enough to
affect $5.5$-m SZE observations in a relatively short integration time.
Observations of these confusing sources were carried out during the period
between 20 November 1993 and 24 January 1994.  Many observations were repeated
several weeks after the initial observations in order to gauge variability.

The desired sensitivity level was achieved in around $400\s$ of integration
time on each source.  On the 40-m telescope, we use the same double
differencing procedure in measuring FLUXes as on the $5.5$-m
(\S~\ref{obs-diff}).  Pointing was checked before each scan on a source.
Calibration was performed by observing DR21, which has a flux of $19.2\pm
0.7\mJy$ at $18.5\GHz$.  Because of the large size of the 40-m telescope,
physical deformation of the dish causes the gain to vary with elevation.  Long
tracks on 3C84 were used to determine the zenith-angle dependent gain
corrections.  We estimate the error for calibration and gain corrections at
$\sim 6\%$, similar to those for the 5.5-m telescope.  

The 40-m $18.5\GHz$ measurements are listed in Table~\ref{tbl:src40m}, along 
with the lower frequency flux density measurements ($S_{low}$) at $4.85\GHz$,
$2.7\GHz$ and $10.7\GHz$ obtained from the literature.  Where the $18.5\GHz$ 
measurement was not significant at the $3\sigma$ level, the $3\sigma$ upper
limit on the flux density is listed.  The deduced spectral
index $\alpha$ for each source, between the lower frequency and $18.5\GHz$, 
are given, where a power-law spectrum is assumed
\begin{equation} 
S \propto \nu^{\alpha}. \label{alpha} 
\end{equation}
The extrapolated flux densities $S_{32}$ are listed in the last column.  These
were calculated using the values for $\alpha$ listed in the table, 
or from upper limits where appropriate.

Because the $18.5\GHz$ frequency of our source measurements is significantly
lower than the 32 GHz observing frequency of our SZE data, and we have no
bracketing measurements at higher frequencies, one should conservatively assume
the $18.5\GHz$ measurements themselves as an upper limit to the 32 GHz source
flux densities.  It is possible that the $18.5\GHz$ emission is dominated by
flat-spectrum compact components in the radio sources.  From the numbers in
Table~\ref{tbl:src40m} we see that this leaves us with an overall factor of two
uncertainty in the source corrections to be applied where a detection at
$18.5\GHz$ was made.  Only better measurements of these sources at frequencies
bracketing $32\GHz$ will allow accurate source subtraction to be made.

\subsection{Corrections to SZE Measurements} \label{src-cor}

The radial distances and position angles (relative to North through East) of
these contaminating sources relative to the field centers are given in
Table~\ref{tbl:src1}.  If the source lies within the reference beam arc
$22\farcm16$ away from the central beam, then the corresponding distance from
the closest arc center $r$ and the parallactic angle $\psi_p$ of the closest
approach are also listed.  The implied equivalent SZE temperatures $\Delta T_a$
are calculated using $6.22\mK\pJy$, since the SZE measurements have already
been converted to main beam temperatures (see equation \ref{jyperk}).  There is
a $6.6\%$ overall conversion uncertainty that includes all of the flux density
scale uncertainties (see \S~\ref{obs-cal}), although this is less than the
uncertainty in the source flux density extrapolation to 32 GHz, and thus has
not been applied to the numbers in Table~\ref{tbl:src1}.  Finally, correction
for the $7\farcm35$ Gaussian beam pattern produces the expected maximum
contributions $\Delta T_{max}$ printed in the last column of the table.

In the MAIN fields, if a source falls within the central part of the main
$5.5$-m beam, it will tend to cancel out the SZE decrement.  Sources in the
reference arcs are {\em subtracted} and thus mimic a decrement, but they will
only be observed at certain parallactic angles.  Such sources are recognizable
by their signature on a plot of the SZE decrement vs. parallactic angle
$\psi_p$.  Sources in the beams of the LEAD and TRAIL fields will have the
opposite effect compared to those in the MAIN field, with levels reduced by a
factor of two due to the averaging (\MLT).  

Because each FLUX measurement is taken at a given parallactic angle, and enters
into the final mean with its individual weight, it is necessary to subtract the
effects of sources in the reference arcs on a point-by-point basis.  In
practice the LEAD, MAIN and TRAIL measurements for each referenced measurement
are adjusted by the values determined using the positions and $\Delta T_a$
from Table~\ref{tbl:src1}.  Only those sources with significant $18.5\GHz$
detections were used.  In Table~\ref{tbl:src5m} we list the unweighted mean
corrections $\Delta T_{fld}$ applied to the data points in the contaminated
fields.  The center beam corrections $\Delta T_{on}$ and maximum reference arc
corrections $\Delta T_{ref}$ (for the parallactic angle where the source is
closest to the center of the arc) are also given.

The corrected SZE results are shown in Table~\ref{tbl:results}.  The second
column lists the source contributions $\Delta T_{\rm 5.5m}(\mbox{src})$
computed using the actual data weighting in the cases of A2142 and A2256, and
the $3\sigma$ limit for A478 (source A478.1).  The corrected values $\Delta
T_{\rm 5.5m}(\mbox{corr})$ are listed in the final column, with the correction
values themselves added in quadrature as an uncertainty.  For A478, the
$1\sigma$ limit on the contribution of source A478.1 was used as the
uncertainty.  The measurement of the SZE in the weakest cluster A2256 is most
adversely affected by the correction uncertainties, with the statistical
standard error increasing from 6.4\% to 12\%.  The error bars on the A478 and
A2142 are not as strongly affected due to the larger relative decrements.
However, in all three cases the corrections applied were similar in magnitude
to or larger than the purely statistical measurement uncertainties, and thus
contribute significantly to the error budgets.  More accurate SZE measurements
will require better source measurements, necessarily contemporaneous with the
SZE observations to deal with possible variability in the foreground sources.

\section{The SZE, Baryonic Mass, and $H_0$} \label{models}

The SZE is proportional to the Compton $y$-parameter
\begin{equation}  \label{y} 
y = \int\limits_{-\infty}^{\infty} \frac{kT_e}{m_e c^2}\sigma_T n_e 
  \, d\zeta \,.
\end{equation}  
For convenience, we will use cylindrical coordinates $(R,\phi,\zeta)$
($R,\zeta$ in Mpc) centered upon the cluster with $\zeta$ along the
line-of-sight.  Then
\begin{equation} \label{yxy}
y(R, \phi) = \int\limits_{-\infty}^{\infty} \frac{k \sigma_T}{m_e c^2}\, 
   n_e( R, \phi, \zeta)\, T_e( R, \phi, \zeta)\, d\zeta \,.
\end{equation}
For the small angles considered here, $R = D_a\,\theta$ for angular diameter
distance $D_a$ to the cluster.  For $q_0=1/2$ and $H_0 = 100\,h\kmsm$, which we
will assume throughout this paper\footnote{At these low redshifts, the effect
of the cosmology is purely kinematic and thus depends only upon $q_0$.  To
first order, $\Delta D_a/D_a \approx \Delta q_0\, z/2$, which at our
redshift limit of $z=0.1$ makes a $\pm 2.5\%$ change in $D_a$, and thus the
derived $h$, for $\Delta q_0 = \pm 1/2$.},
\begin{equation}\label{da}
 D_a = 6000\,\frac{(1+z) - \sqrt{1+z}}{(1+z)^2}\,\hmpc.
\end{equation}

The on-sky intensity differences are measured in units
of {\it antenna temperature} $\Delta T_a$, which is the equivalent
temperature difference in the Rayleigh-Jeans limit  
\begin{equation} \label{ta}
\Delta I_{\nu} = \frac{2k\nu^2}{c^2}\,\Delta T_a.
\end{equation}
Using the standard formulae for the fractional change in the intensity 
of the thermal background in the non-relativistic limit (Sunyaev \& Zeldovich 
\markcite{sz80} 1980), we get the frequency dependence of the measured change
in antenna temperature of the microwave background due to the SZE:
\begin{equation} \label{szd}
\frac{\Delta T_a}{T_{cmb}} = y\frac{x^2 e^x}{(e^x - 1)^2}\,
   \left(x \coth{\frac{x}{2}} - 4\right),
\end{equation}
where $x$ is the dimensionless frequency
\begin{equation} \label{xfrq}
x = \frac{h\nu}{kT_{cmb}} = \frac{\nu}{56.80\,{\rm GHz}}.  
\end{equation}
We use the COBE FIRAS value for the microwave background temperature $T_{cmb} =
2.726\pm0.010$~K (Mather \etal \markcite{mather94} 1994).  At the $5.5$-m 
observing frequency $\nu=32\GHz$ ($x=0.563$) this is
\begin{equation} \label{y32}
\frac{\Delta T_a}{T_{cmb}} = -1.897\,y \,.
\end{equation} 

There are several possible corrections to the expressions (\ref{szd}) and
hence (\ref{y32}).  In addition to the thermal SZE, there is a kinematic
effect due to the peculiar velocity of the cluster (Sunyaev and Zeldovich
1980).  At 32~GHz, a peculiar velocity of $300\kms$ for a cluster with
$kT_e=7.5\keV$ will produce a change in the SZE intensity of only 2\%, and can
safely be ignored in these calculations.

Another factor not accounted for in our expression for $y$ is the
relativistic correction to (\ref{szd}), which was derived in the
non-relativistic limit.  Rephaeli \markcite{reph95} (1995) has calculated the
corrections for the low optical depths ($\tau < 10^{-2}$) and mildly
relativistic electron temperatures ($kT_e \sim 5$ -- $10 \keV$) appropriate to
our clusters.  For these parameters, Rephaeli finds corrections of around
$+3\% \pm 0.3\%$ at our observing $x=0.563$, \ie the SZE decrement is less 
pronounced in magnitude than what eq. (\ref{szd}) would predict.  Because 
this is a systematic underestimation of $y$ given an observed $\Delta T_a$, we
use the corrected relation 
\begin{equation} \label{y32corr}
y_{meas} = -\,\frac{\chi_{rel}\,\Delta T_a}{1.897\,T_{cmb}}
\end{equation} 
where the relativistic correction factors $\chi_{rel}$ for each cluster are
given in Table~\ref{tbl:relcor}, along with the Compton parameters $y_{meas}$
using the source-corrected $\Delta T_a$ from Table~\ref{tbl:results}.

Herbig \etal \markcite{herbig95} (1995) carried out SZE observations and
analysis of the Coma cluster using the same $5.5$-m setup and calibration 
scale as was employed in our observations.  With source corrections
and calibration uncertainty included, they found a non-relativistic 
$y$-parameter of $y_{meas} = (5.96\pm0.99)\times10^{-5}$.
If we apply the relativistic correction $1.029$ to the 
Herbig \etal measurement, and remove the $6.9\%$ calibration uncertainty,
we get $y_{meas} = (6.13\pm0.93)\times10^{-5}$.  This is the value we
have given in Table~\ref{tbl:relcor}.

The observed SZE decrement is the true decrement modified by the telescope
primary beam and the beam switching.  The 5.5-m single beam pattern is well
approximated by a circular Gaussian
\begin{equation} \label{gbeam}
g(\theta) = \frac{1}{2\pi \theta_g^2}\,
 \exp \left( - { \theta^2 \over 2\,\theta^2_g } \right) 
\end{equation}
with beam width $\theta_g = 3\farcm12 \pm 0\farcm04$ ($7\farcm35 
\pm 0\farcm10$ FWHM).  The beam is less than
$1.4\%$ elliptical.  Because clusters track through a range of parallactic
angles, the slightly elliptical beam is rotated on-sky and therefore the
effective beam is well represented by the geometric mean $\theta_g$.

The average Compton $y$-parameter in the Gaussian beam, on a line of sight
offset at cylindrical radius $R$ from the center at position angle $\phi$, 
for small angles is given by 
\begin{equation} \label{yg}
y_g(R,\phi) = \int\limits_{0}^{2\pi} d\theta \int\limits_0^{\infty}
   r\, dr \frac{1}{2\pi L_g^2} e^{- d^2/ 2 L_g^2} y(r, \phi-\theta) 
   \quad\qquad d^2 = R^2 + r^2 - 2 R r \cos(\theta).
\end{equation}
Here $L_g = D_a\,\theta_g$.  The beam-switching can be evaluated as
\begin{equation} \label{yswitch}
y_{sw}(R,\phi;\psi_p) = y_g(R,\phi) - {1\over2} y_g(R_{-},\phi_{-}) - 
   {1\over2} y_g(R_{+},\phi_{+})
\end{equation}
with
\begin{mathletters}
\begin{eqnarray}
   R_{\pm}^2 & = & D^2\sin^2(\psi_p + \phi) + \left[ R \pm
     D\cos(\psi_p + \phi) \right]^2 \\
   \tan \phi_{\pm} & = & \frac{R\sin\phi \mp D\sin\psi_p}
     {R\cos\phi \pm D\cos\psi_p}.
\end{eqnarray}
\end{mathletters}
The 5.5-m beam separation is $\theta_D=22\farcm16$, so $D = D_a\theta_D$, 
and $\psi_p$ is the principal parallactic angle for the observation. 
For convenience, the position angle $\phi$ is measured starting from the
East so it is in the same orientation as the parallactic angle $\psi$.

We will most often use the integrals through the cluster center in a
cylindrically symmetric model.  The angular dependences are dropped,
so $y_g(R,\phi)=y_g(R)$ is now a function of angular radius from the cluster center only.  The expressions (\ref{yg}) and (\ref{yswitch}) are abbreviated as
\begin{equation} \label{ygcenter}
y_g \equiv y_g(0) = 2\pi \int\limits_0^{\infty}
   r\, dr \frac{1}{2\pi L_g^2} e^{- r^2/ 2 L_g^2} y(r).
\end{equation}
and 
\begin{equation} \label{yocenter}
y_{sw} \equiv y_{sw}(0) = y_g(0) - y_g(D).
\end{equation}

\subsection{Density models for the ICM} \label{models-dens}

The distribution of the X-ray emitting gas in galaxy clusters has frequently
been modeled by an isothermal $\beta$-model (Cavaliere \& Fusco-Femiano 
\markcite{cavaliere76} 1976), also known as
a modified isothermal King model.  In this spherically symmetric
model, the electron gas density $n_e$ is given as a function of the
spherical radius $r$
from the center of the cluster by
\begin{equation} \label{king}
n_e = n_0 \left(1 + \frac{r^2}{r_c^2}\right)^{-3 \beta /2}, 
\end{equation} 
where $r_c$ is the core radius of the cluster and $n_0$ is the density at
$r=0$.  For the present, we will consider mass distributions with circular
symmetry in the plane of the sky, although it is easy to generalize to
ellipsoidal profiles.   

Table~\ref{tbl:xray} lists published parameters of the intracluster gas from
various X-ray observations of A478, A2142, A2256, and Coma.  The core radius
$\theta_{core}$ determined from the X-ray surface brightness profile is
listed, and has been converted into a linear size $r_c$ in $\hmpc$ using
(\ref{da}).  The overall temperature of the X-ray emitting gas $T_e$ is given
as $k T_e$ in keV ($10^8\Kel = 8.61\keV$).  As determined from the X-ray
emission, the central densities $n_0$ are in units of $\hcc$.  A value for
$\beta$ is given only if it has been determined by fitting the surface
brightness profile.  All errors are given as $1\sigma$, converted from 90\%
confidence ($\approx 1.645\sigma$) in the literature if necessary.

We now discuss the models for each cluster in detail.  Because the
model uncertainties are in most cases the dominant source of systematic
error in the determinations of the baryonic masses and Hubble constant
for this sample, we plan to make our own detailed analysis of the \rosat
data for these clusters to obtain more accurate models, and to understand
the limitations of our particular method better.  For now, we adopt the models 
presented in the literature and proceed with our analysis.

\subsubsection{A478}

Abell 478 has been shown to contain one of the largest cluster cooling flows
with more than $5\times10^{11}\hmsun$ of X-ray absorbing matter within the
inner $150\hkpc$ and a total mass deposition rate of $\sim 500 \hmsun {\rm
yr^{-1}}$ (Johnstone \etal \markcite{johns92} 1992, Allen \etal
\markcite{allen93} 1993).  The combined \ginga and \rosat data measure the
temperature $kT_e = 6.56\pm0.09\keV$ for the cluster isothermal component, and
a temperature of $kT_e \sim 3\keV$ within the inner $75\hkpc$.  The ROSAT PSPC
image of A478 (Allen \etal \markcite{allen93} 1993) shows an axial ratio of
$\sim 0.8$ to the inner ($\theta<2\farcm4$) isophotal contours.  This would
suggest that A478 is probably even more ellipsoidal than this, and should
be kept in mind in the proceeding analysis.

The presence of such a large cooling component to the cluster core medium
makes modeling of the SZE from the X-ray emission difficult.  Edge \& Stewart
\markcite{edge91a} (1991a) found $T_e=6.8\keV$, $r_c=0.10\hmpc$ and
$n_0=25.2\pm2.8\,\times10^{-3}\hcc$ from \exosat observations.  Allen \etal
\markcite{allen93} (1993) fit a central electron density to the isothermal
(non-cooling) component of $n_0 \approx 9.55\times10^{-3}\hcc$ assuming a core
radius of $r_c=0.125\hmpc$ and a King profile ($\beta=2/3$).

We have adopted the Allen \etal \markcite{allen93} (1993) values for $n_0$,
$r_c$ and $\beta$.  The uncertainties on these quantities were determined
empirically by comparison with our own preliminary analysis of the \rosat
data.  This model is the one listed in Table~\ref{tbl:xray}.

\subsubsection{A2142}

A2142 is the most distant cluster ($z=0.0899$) and has the largest 2-10 keV
luminosity in our sample.  A2142 is also the second most luminous cluster in
the Edge sample as a whole.  Edge \& Stewart \markcite{edge91a} (1991a) list
A2142 as a cooling core cluster, and Edge, Stewart \& Fabian \markcite{edge92}
(1992) derive a mass flow rate of 50--$150\hmsun\,{\rm yr^{-1}}$.

In the compilation of cluster temperatures by David \etal \markcite{david93}
(1993), A2142 is listed as having a {\ginga} temperature of
$kT_e=8.68\pm0.12\keV$.  Abramopolous \& Ku \markcite{abramku83} (1983) derive
$r_c=0.26\pm0.01\hmpc$ and $n_0=6.97\pm0.41\expo{-3}\hcc$, where we have
estimated the uncertainty in $n_0$ through the relation $n_0 \propto
L_x^{1/2}\,r_c^{-3/2}$.  The parameter $\beta$ is fixed at unity in this
model.  For want of a better determination, we adopt $\beta=1\pm0.3$.

\subsubsection{A2256}

Although A2256 does not appear to have a central cooling flow (Edge, Stewart
\& Fabian \markcite{edge92} 1992), it does show significant substructure in
the X-ray emitting gas.  Briel \etal \markcite{briel91} (1991) found evidence
for a ``merger event'' in the \rosat PSPC image of the cluster.  Two surface
brightness peaks were found in the cluster center with a separation of
$3\farcm5$ ($160\hkpc$), with some indication of differing temperatures.  David
\etal \markcite{david93} (1993) list an overall \ginga temperature $kT_e =
7.51\pm0.11\keV$, while the \rosat PSPC data gives $kT_e \sim 2.0\keV$ for the
cooler (NW) subcluster.  Fits of the azimuthally averaged data excluding the
secondary maximum to a modified isothermal King profile yielded
$\theta_{c,1}=4\farcm83\pm0\farcm17$ and $\beta_1=0.756\pm0.013$, while a fit
to the secondary after subtraction of the primary smooth profile gave
$\theta_{c,2}=4\farcm3\pm0\farcm4$ and $\beta_2=1.1\pm0.1$ and a peak surface
brightness 82\% of that of the primary.  Briel \etal \markcite{briel91} (1991)
also analyze the distribution of 87 galaxies to find radial velocity
dispersions of $1270\pm127 \kms$ for the main cluster region and
$250\pm123\kms$ in the NW subcluster.  A relative systemic velocity difference
of $-2150\pm259\kms$ is found between the NW region and the main cluster ---
this difference is consistent with the infall velocity at 1 Mpc from a
$10^{15}\msun$ cluster.  The pointing center used in our observations is
approximately in the center, between the two components of A2256.

Davis \& Mushotzky \markcite{davmush93} (1993) examined \eins IPC data and
also found evidence for the merger, and derive $\theta_c=6\farcm0{ +0.9 \atop
-0.7 }$ and $\beta=0.72{ +0.10 \atop -0.08}$.  By fitting elliptical
isophotes, we find axial ratios of $0.6$ near the center to $0.25$ at a radius
of $10\arcm$.  Spectroscopy from the {\it BBXRT} (Miyaji \etal
\markcite{miyaji93} 1993) indicates a temperature of $kT_e=4.6{+0.9 \atop
-0.7}\keV$ for the NW component.

A more detailed analysis of the A2256 \rosat PSPC data has been carried out by
Henry, Briel, \& Nulsen \markcite{henry93} (1993).  From their data, an
isothermal model of the intra-cluster medium has been derived:
$\theta_c=5\farcm33\pm0\farcm20$, $\beta=0.795\pm0.020$,
$n_0=3.55\pm0.18\expo{-3}\hcc$.  They derived a low-energy temperature from
the \rosat data of $kT_e=6.9\pm0.6 \keV$, which is consistent with the \ginga
temperature.  We choose to adopt the Henry \etal\ parameters and the
David \etal \ginga temperature.  

\subsubsection{Coma}

The Coma cluster is also a member of our sample.  Briel, Henry \& B\"{o}hringer
\markcite{briel92} (1992) present an isothermal X-ray model for the Coma gas
with $\beta=0.75\pm0.03$ and $\theta_c=10\farcm5\pm0\farcm6$
($r_c=0.207\pm0.012\hmpc$ for an assumed redshift of $z=0.0235$).  They adopt
an electron temperature of $kT_{eff}=8.2\pm0.2\keV$, obtained from \ginga
measurements.

Earlier observations by Hughes, Gorenstein \& Fabricant \markcite{hughes88}
(1988) with \exosat give a slightly higher temperature of
$kT_{eff}=8.5\pm0.3\keV$.  After deprojection and subtraction of galactic
absorption, they fit a model with $\beta=0.63\pm0.03$ and
$\theta_c=7\farcm6\pm0\farcm4$, and central electron density of 
$n_0 \approx 3\expo{-3}\hcc$.  For a best-fit model using the 
\exosat and \tenma data, they assume an
isothermal core with the high temperature of $kT_{iso}=9.1\pm0.4\keV$,
which extends out to a radius of $\theta_{iso}=23\arcm{+12\arcm \atop
-8\arcm}$, beyond which the temperature falls almost adiabatically (polytropic
index $\gamma \sim 1.555$).  This gives a temperature profile outside
the isothermal radius $R_{iso}$ of
\begin{equation} \label{tpoly}
T(R) = T_{iso}\,\left[ { 1 + (R/r_c)^2 \over 1 + (R_{iso}/r_c)^2 } \right]
   ^{-{3\beta \over 2}(\gamma - 1)} \qquad\qquad R > R_{iso}.
\end{equation} 
This is the model adopted by Herbig \etal \markcite{herbig95} (1995), 
although with $n_0$, $r_c$ and $\beta$ as given by Briel, Henry \& 
B\"{o}hringer.

The higher temperature may in fact be indicated by \asca observations of Coma
(see Fabian \etal \markcite{fabian94} 1994) which prefer temperatures of
$kT_{eff} \sim 9\keV$.  For consistency with Herbig \etal \markcite{herbig95}
we adopt the Briel, Henry \& B\"{o}hringer $n_0$, $r_c$ and $\beta$, and a
temperature $kT_{eff}=9.1\pm0.4\keV$.  In the model calculations in this
paper, we do not include an isothermal cutoff, and this makes only a
few percent difference.  

\subsubsection{Model-Dependent Quantities}

Using the isothermal $\beta$-model with electron temperature $T_e$, the
$y$ at a point at projected radius $t=R/r_c$ from the cluster center
(\ref{y}) has the familiar form 
\begin{equation} \label{yt}
  y(t) = \frac{k T_e}{m_e c^2}\,n_0\,\sigma_T\, r_c \,
  \int_{-\infty}^{\infty} ds\,
  \left(1 + t^2 + s^2 \right)^{- \frac{3\beta}{2}}. 
  = y_0\,\left(1 + t^2 \right)^{\frac{1}{2} - \frac{3\beta}{2}} 
\end{equation} 
where $y_0$ is the $y$-parameter at zero projected radius
\begin{equation} \label{dt}
  y_0 = 7.12\times10^{-5} \, h^{-1/2}\,
  \frac{\Gamma\left(\frac{3\beta-1}{2}
  \right)}{\Gamma \left(\frac{3\beta}{2} \right)}\, \left (\frac{n_0}{10^{-3}
  \hcc} \right) \left( \frac{T_e}{10\keV} \right) \left(\frac{r_c}{\hmpc}
  \right).
\end{equation}  
The dependence on $h$ is due to the choice of the units for $n_0$ and
$r_c$, which in turn are determined from the X-ray measurements.  The
resulting factor of $h^{1/2}$ will be used to determine the value of the 
Hubble constant by comparison with the observed $y$-parameters.

Given the model for the density profile $n_e( R, \phi, \zeta)$ in the cluster,
we can determine the efficiency $\eta_{obs}$ at which our switched observations
can recover the SZE that an ideal pencil-beam through the cluster center would
measure
\begin{equation} \label{etaobs}
 \eta_{obs} = \frac{y_{sw}}{y_0}.
\end{equation} 
In addition, we can compute the efficiency $\eta_g$ at which 
the SZE is measured with respect to an ideal Gaussian main beam 
\begin{equation} \label{etag}
 \eta_{g} = \frac{y_{sw}}{y_g}\,.
\end{equation} 
These efficiencies depend upon the model only through $\theta_c$ and $\beta$.
The derived efficiencies (for a pointing center at the cluster center $R=0$)
for the OVRO 5.5-m SZE observations are given in Table~\ref{tbl:relcor}.  The
uncertainties in the $\eta_{obs}$ and $\eta_{g}$ were determined numerically 
using the stated uncertainties in the model $\theta_c$ and $\beta$.

Other model-derived quantities of interest are the equivalent spherical volume
\begin{equation} \label{sphvol}
V_s(R) = 4\,\pi\,r_c^3\,\int_0^{R/r_c} dt\,t^2\,(1+t^2)^{-3\beta/2}
\end{equation} 
and the Gaussian cylindrical volume
\begin{equation} \label{gauvol}
V_g(L_g) = 2\,\pi\,r_c^3\,\frac{\Gamma\left(\frac{1}{2}\right)\,
  \Gamma\left(\frac{3\beta-1}{2}
  \right)}{\Gamma \left(\frac{3\beta}{2} \right)}
  \,\int_0^{\infty} dt\,t\,\exp(-\frac{r_c^2 t^2}{2 L_g^2})
  ( 1 + t^2 )^{1/2 - 3\beta/2}.
\end{equation} 
These quantities are the equivalent volumes for a uniform density cluster at
the central density $n_0$\footnote{Note that equations (\ref{sphvol}) and
(\ref{gauvol}) can respectively be written in terms of the Incomplete Beta
Function and the Incomplete Gamma Function.  However, it is easiest to
evaluate these integrals numerically, using Maple or Mathematica for example.
}.  These volumes are important for
relating the observed X-ray emission and the observed SZE to the implied
baryonic mass contained within the cluster.

\subsection{Baryonic Mass in Clusters}

Given knowledge about the electron temperature in the intracluster gas,
we can use the $y$ parameter to measure a baryonic mass $M_b$ for the ionized
phase.  For a general density model
\begin{equation} \label{massb}
 M_b = \int \int \int dR\, d\phi\, d\zeta\, 
  \alpha^\prime m_b\, n_e( R, \phi, \zeta).
\end{equation}
Comparison with equation (\ref{y}) gives for our cylindrical model
\begin{equation}  \label{massb2}
 M_b = \frac{m_e c^2}{\sigma_T\, k T_{eff}} \alpha^\prime m_b\, 2\pi\, 
  \int dR\, y(R),
\end{equation}
where $\alpha^\prime$ is the number of baryons per electron, and $m_b$ is the
baryon (nucleon) mass.  Considering only H and He at 12:1 in number of atoms,
$\alpha^\prime \approx 8/7$.  For a general temperature distribution,
the effective temperature is given by
\begin{equation} \label{teff}
 T_{eff} = \frac{\int \int \int  dR\, d\phi\, d\zeta\, 
   n_e( R, \phi, \zeta)\, T_e( R, \phi, \zeta)}
   {\int \int \int  dR\, d\phi\, d\zeta\, n_e( R, \phi, \zeta) }
\end{equation}
which reduces to the single temperature $T_{eff} = T_e$ for an isothermal 
cluster medium at electron temperature $T_e$.

By combining (\ref{massb2}) with (\ref{ygcenter}), we can determine the 
temperature
weighted baryonic mass within the cylinder defined by the $7\farcm35$ FWHM
Gaussian beam of the 5.5-m telescope through the cluster center
\begin{equation} \label{mgas}
 M_g  = \alpha^\prime m_b\,n_0\,V_g =
 \frac{m_e c^2}{\sigma_T\, k T_{eff}}\, 2\pi L^2_g \,\alpha^\prime m_b\, y_g\ ,
\end{equation}
where $L_g=D_a\,\theta_g$ as before.  Hence, the baryonic mass may be
written as 
\begin{equation} \label{mbary}
 M_g = 4.407\times10^{14}\, \bigg(\frac{1\,\hbox{keV}}{T_{eff}}\bigg)
   \bigg(\frac{L_g}{h^{-1}\,\hbox{Mpc}}\bigg)^2
   \bigg(\frac{ y_{sw} }{ 10^{-5}\,\eta_g }\bigg)
   \,h^{-2}\,M_\odot.
\end{equation}
The efficiency factor $\eta_g = y_{sw}/y_g$ converts the measured $y_{sw}$
into the $y_g$ within the Gaussian main beam.  A better representation of the
SZE in terms of a mass is the surface baryonic mass density within the Gaussian
cylinder
\begin{equation} \label{sgas}
 \Sigma_g = \frac{M_g}{2\pi L^2_g}
   = 7.013\times10^{13}\, \bigg(\frac{1\,\hbox{keV}}{T_{eff}}\bigg)
   \bigg(\frac{ y_{sw} }{ 10^{-5}\,\eta_g }\bigg)
   \,M_\odot\,{\rm Mpc}^{-2}.
\end{equation}
The surface density is distance independent, and is a more consistent
parameter than the mass, which will vary with the resolution $L_g$.  The
values to use for $y_{sw}$ are the measured $y$-parameters $y_{meas}$, with
relativistic corrections applied using (\ref{y32corr}), found in
Table~\ref{tbl:relcor}.  The derived baryonic masses for our clusters are
given in Table~\ref{tbl:bmass}.  The clusters have similar surface densities
$\Sigma_g \sim 7 \times 10^{13} M_\odot\,{\rm Mpc}^{-2}$.

The factor $\eta_g$ and the effective temperature $T_{eff}$ are the only
model-dependent quantities in $M_g$ and $\Sigma_g$.  In Table~\ref{tbl:bmass},
we use the X-ray model parameters listed in Table~\ref{tbl:xray}.  Since
(\ref{mbary}) and (\ref{sgas}) are linear in the observable $y_{sw}$, the SZE
is potentially a more accurate probe of the baryonic mass than the X-ray
emission.  

The X-ray emission from clusters has been used to determine the baryonic mass
fraction by comparison with derived total masses.  This calculation has been
done for Coma (White \etal \markcite{white93} 1993) and A2256 (Henry, Briel,
\& Nulsen \markcite{henry93} 1993).  The SZE is an independent measure of 
the mass within the Gaussian cylinder of the beam.  We can use the X-ray
derived model to relate $M_g$ to the mass within the sphere
\begin{equation}
M_{sze}(R) = M_g\,\frac{V_s(R)}{V_g}
\end{equation}
where $V_g$ is the Gaussian volume (\ref{gauvol}) within the beam on the
cluster.

For our clusters A478, A2142, and A2256, as well as Coma, the baryonic masses
are given in Table~\ref{tbl:bmass}.  In Table~\ref{tbl:bfrac}, gravitational
masses have been obtained from the literature and the baryonic fraction within
some given fiducial radius $R_0$ is computed.  The model-dependent factors
$V_s/V_g$ are listed for the assumed $R_0$, along with the 1-$\sigma$
uncertainties computed from the model uncertainties in $r_c$ and $\beta$.

White \& Fabian \markcite{white95} (1995) discuss the ``baryon overdensity''
problem in the context of \eins observations of a number of clusters,
including A478 and A2142.  Henry, Briel, \& Nulsen \markcite{henry93} (1993)
give detailed models and masses for A2256, and White \etal \markcite{white93}
(1993) compute the enclosed baryonic and gravitational masses for Coma.  We
discuss the results for each cluster, and the four clusters as a group, below.

\subsubsection{A478} 

White \& Fabian \markcite{white95} (1995) consider a radius of $R_0=0.976\hmpc$
within which they find an X-ray determined gas fraction
$M_{xray}/M_{tot}=0.091\pm0.008\,h^{-3/2}$.  No uncertainties in the values for
$M_{tot}$ are stated, although they are likely to be high (probably 20\% or
more).  This should be kept in mind when evaluating the uncertainties for the
clusters listed in the White \& Fabian paper.

For the model in Table~\ref{tbl:xray}, our SZE measurements give a 
Gaussian mass of $M_g=(2.58\pm0.21)\times10^{13}\,h^{-1}\,M_\odot$ within the 
5-m beam.  This model gives a ratio $V_s/V_g = 2.99\pm0.08$ 
within $R_0=0.976\hmpc$, so we find an SZE-indicated baryonic mass 
of $M_{sze}=(7.71\pm0.66)\times10^{13}\,h^{-2}\,M_\odot$.  Using the
gravitational mass from White \& Fabian, we get a baryonic fraction
of $M_{sze}/M_{tot} = 0.166\pm0.014\,h^{-1}$. 

The cluster A478 stands out with a higher baryonic fraction from both the X-ray
and SZE measurements, and has a stronger SZE decrement than expected from the
X-ray measurements (giving a lower implied Hubble constant from the ratio
$M_{xray}/M_{sze}$), when compared with the other clusters in this sample (see
below, and in the next section).  These discrepancies may be explained by
elongation of the cluster along the line of sight, as indicated by its observed
ellipticity in the plane of the sky.  We will discuss this further in the
context of the Hubble constant in the next section.

\subsubsection{A2142}

For A2142, White \& Fabian \markcite{white95} (1995) find an X-ray determined
gas mass fraction $M_{xray}/M_{tot}=0.050\pm0.003\,h^{-3/2}$ within
$R_0=0.976\hmpc$.  Our SZE measurements give a Gaussian mass of
$M_g=(2.08\pm0.22)\times10^{13}\,h^{-1}\,M_\odot$ with an efficiency of
$V_s/V_g = 2.90\pm0.43$ for $R_0=0.976\hmpc$.  Therefore,
$M_{sze}=(6.03\pm1.10)\times10^{13}\,h^{-2}
\,M_\odot$ and $M_{sze}/M_{tot} = 0.060\pm0.011\,h^{-1}$. 

\subsubsection{A2256}

Henry, Briel, \& Nulsen \markcite{henry93} (1993) fit a model with
$r_c=0.245\pm0.009\hmpc$, compared to the 5-m beam size of $L_g=0.143\hmpc$ at
redshift $z=0.0581$.  They derive an enclosed mass of
$M_{tot}=(5.1\pm1.4)\times10^{14}\,h^{-1}\,M_\odot$ within $R_0=0.76\hmpc$.
From the X-ray data they find $M_{xray}/M_{tot}=0.063\pm0.039\,h^{-3/2}$.
Assuming our isothermal model, we find a ratio $V_s/V_g = 4.29\pm0.05$ within
the sphere of radius $R_0$.  The SZE measurements gave a Gaussian mass of
$M_g=(7.1\pm0.9)\times10^{12}\,h^{-1}\,M_\odot$ within the 5-m beam, thus
$M_{sze}=(3.05\pm0.39)\times10^{13}\,h^{-2}\,M_\odot$ in the sphere, and 
$M_{sze}/M_{tot} = 0.060\pm0.018\,h^{-1}$.  

\subsubsection{Coma}

For Coma, White \etal \markcite{white93} (1993) adopt a (model-dependent)
total mass of $M_{tot} = (1.10\pm0.22)\times10^{15}\,h^{-1}\,M_\odot$ within
a sphere of radius $R_0=1.5\hmpc$ (the Abell radius).  They find 
$M_{xray}/M_{tot} = 0.050\pm0.013\,h^{-3/2}$.
Our adopted isothermal model with an assumed temperature of 
$kT_{eff}=9.1\pm0.4\keV$ gives the Gaussian
mass of $M_g=(1.81\pm0.30)\times10^{12}\,h^{-1}\,M_\odot$ within the 5-m beam.
For a spherical radius of $R_0=1.5\hmpc$, the ratio $V_s/V_g=38.3\pm3.0$,
and thus $M_{sze} = (6.93\pm1.27)\times10^{13}\,h^{-2}\,M_\odot$.  Using this
value, we find $M_{sze}/M_{tot} = 0.063\pm0.017\,h^{-1}$, in agreement with
the White \etal\ fraction for $h = 0.62{+0.36 \atop -0.28}$.  

For the Coma cluster, the $7\farcm35$ 5-m beam is small
($L_g=61.5\hkpc$) compared to the Abell radius $R_0=1.5\hmpc$, and we are
making a large Gaussian correction $V_s/V_g \sim 38$.  In addition, the
contribution of the SZE signal in the reference beams is significant, so the
details of the electron temperature profile in these outer parts 
are more important than in the other clusters.

\subsubsection{The Sample}

We see that in three of the four clusters (A2142, A2256, and Coma) the SZE
determined baryonic fractions $M_{sze}/M_{tot}$ are consistent, with a mean of
$0.061\pm0.010\,h^{-1}$ (the uncertainty from the individual error bars, not
the scatter).  We exclude A478 from this average due to the discrepancies
between this cluster and the others in the sample (if A478 is included the
mean becomes $0.087\pm0.030\,h^{-1}$).  We should also include the $6.9\%$
calibration uncertainty, giving $\langle M_{sze}/M_{tot} \rangle =
0.061\pm0.011\,h^{-1}$.

Strictly speaking, this is a {\it lower limit} on the baryon fraction
$M_B/M_{tot}$, since we have not included the luminous mass in galaxies, and
some of the dark matter may be baryonic.  White \etal \markcite{white93}
(1993) find a ratio $M_{gal}/M_{tot} = 0.009 \pm 0.003$ in Coma, compared to
the fraction $M_{xray}/M_{tot} = 0.050\pm0.013\,h^{-3/2}$ in hot gas.  Henry,
Briel, \& Nulsen \markcite{henry93} (1993) find similar relative fractions in
galaxies and gas for A2256.  Thus, we can safely assume that the luminous
galaxies contribute around 20\% or less of that mass contributed by the hot
IGM.  If we use the Coma value, and apply it to the sample as a whole, then
$\langle M_B/M_{tot} = 0.009\pm0.003 + 0.061\pm0.011\,h^{-1}$.  Note that
for the low values of $h$ which are generally preferred, the contribution 
from luminous galaxies is further reduced relative to the gas.  In this paper,
we will use the SZE mass as a lower limit on the total baryon mass.

Standard estimates of the fraction of closure density in baryons for
homogeneous big-bang nucleosynthesis give $0.011 \leq \Omega_B h^2 \leq 0.015$
($2\sigma$) (Smith, Kawano \& Malaney \markcite{smith93} 1993).  However,
recent measurements of the deuterium abundance in the Lyman-$\alpha$ forest
clouds in several QSOs lead to incompatible values for $\Omega_B$ that lie
outside this range.  One group finds a low value for the deuterium abundance
(Tytler, Fan and Burles \markcite{tytler96} 1996) implying a high $\Omega_B
h^2 = 0.024\pm0.006$, while the other finds a high deuterium abundance (Rugers
\& Hogan \markcite{hogan96} 1996) implying a low $\Omega_B h^2 = 0.0062 \pm
0.0008$.

We can now estimate the total mass density parameter
\begin{equation}
\Omega_0 = \frac{\Omega_B}{M_B/M_{tot}} \leq \frac{\Omega_B}{M_{sze}/M_{tot}}
\end{equation}
assuming that the baryonic fraction in clusters reflects that of the Universe
as a whole.  Using our reduced sample average $\langle M_{sze}/M_{tot} \rangle
= 0.061\pm0.011\,h^{-1}$, and assuming standard nucleosynthesis limits
$\Omega_B h^2 = 0.013\pm0.002$, one obtains $\Omega_0 h \leq 0.21\pm0.05$ (and
$\Omega_0 h \leq 0.15\pm0.06$ if A478 is included in the mean).  However, if
one adopts the higher $\Omega_B h^2 = 0.024\pm0.006$, then our data imply a
significantly higher limit $\Omega_0 h \leq 0.39\pm0.12$.  On the other hand,
if we use the high-deuterium value giving a low $\Omega_B h^2 = 0.0062 \pm
0.0008$, then we find a low density parameter $\Omega_0 h \leq 0.10\pm0.02$.

Using the standard nucleosynthesis values for $\Omega_B$, we find the cluster
data is consistent with $\Omega_0=1$ only for very low values of the Hubble
constant ($h\approx0.21$), or for similar values of baryonic mass segregation
($\Omega_{B,tot}/\Omega_{B,clus} \approx 0.21$).  Neither of these is
indicated by other cosmological data.  Extremely low values of $h$ are not
consistent with the estimates derived by comparison with the X-ray emission
except in the case of A478 (see below).  On the other hand, this result is
consistent with large-scale structure studies which yield values in the range
$0.2 \lesssim \Omega_0 h \lesssim 0.3$ (Efstathiou, Sutherland \& Maddox
\markcite{apm90} 1990, Efstathiou, Bond \& White \markcite{ebw92} 1992,
Peacock \& Dodds \markcite{peacock94} 1994).  This widespread ``baryon
overdensity'' problem has been seen in many clusters (e.g. White \& Fabian
\markcite{white95} 1995).

However, if the higher $\Omega_B$ from a low deuterium abundance is correct,
then the values of the Hubble constant implied by comparison with the X-ray
data (see below) would allow a critical density for the Universe.  Conversely,
adoption of the high deuterium, low baryon density $\Omega_B$ would only
exacerbate the baryon overdensity problem, forcing us to accept a low density
universe.

Departures from isothermality or the coexistence of multiple phases in the
intracluster medium will introduce errors in our determination, as will model
errors in the extrapolation to the spherical masses.  In particular, the
determined values for the total binding masses are uncertain.  Measurements
of total mass surface density from weak gravitational lensing would be 
particularly well-suited to this method, as the angular size of a typical
CCD frame is similar to that of the $5.5$-m beam width.

Note that our SZE (and X-ray) measurements only count the baryons in the
hot IGM.  The luminous baryonic matter in galaxies, and any non-luminous
baryonic matter (such as in brown dwarfs, Jupiters, or compact objects)
would be in addition to this estimate.  Thus, we place a {\it lower limit}
on the total baryonic mass, and therfore the baryonic fraction.  The
baryon overdensity problem would only get worse if there were substantial
contributions from these other baryon resevoirs.  

The difference between the SZE-based and X-ray based estimates of the baryonic
fraction is due to the different dependences on the Hubble constant, $h^{-1}$
versus $h^{-3/2}$ respectively.  If we compare the X-ray and SZE numbers given
above for A2142, A2256 and Coma (thereby excluding A478), the average ratio is
$\langle M_{xray}/M_{sze} \rangle = 0.889\pm0.099\,h^{-1/2}$, and therefore in
agreement for $h = 0.79{+0.19 \atop -0.17}$.  We explore this in more detail in
the next subsection.

\subsection{The Hubble Constant}

The key to the determination of $H_0$ lies in the observation that the X-ray
brightness and SZE decrement scale differently with temperature and density.
Because the core radius $r_c$ is determined from the observed angular size of
the cluster ($r_c \propto h^{-1}$), the central densities of the cluster gas
determined from X-ray data are proportional to $h^{1/2}$.  The angular
diameter--distance relation introduces a factor of $h^{-1}$ into the
$y$-parameter for the dependence on $r_c$ (see equation \ref{dt}).  Thus,
$y_{pred}$, the estimate of the switched measurements of the Compton $y_{sw}$
derived from the X-ray model in Table~\ref{tbl:xray} using (\ref{yocenter})
and (\ref{ygcenter}), are proportional to $h^{-1/2}$, and the observations of
the actual SZE $y_{meas}$ can therefore be used to find $H_0$:
\begin{equation} \label{h} 
h = \left(\frac{y_{pred}}{y_{meas}}\right)^2 .
\end{equation}
The SZE model predictions $y_{pred}$, the measured $y_{meas}$, and the inferred
Hubble constant values are given in Table~\ref{tbl:hubble}.  The uncertainties
in $y_{pred}$ were computed from the model parameter uncertainties.  These
parameters were assumed to vary independently, though in fact they are
correlated from the X-ray fitting procedure (particularly $\theta_c$ and
$\beta$).  A more direct approach, comparing the X-ray data and SZE data
directly, as in Birkinshaw \etal \markcite{birk91} (1991) and Birkinshaw \&
Hughes \markcite{birk94} (1994), would be preferable.  As it is, using the
available information, the uncertainties quoted here are likely to be slightly
inflated, as the parameter correlations will reduce the overall uncertainty
somewhat.  The asymmetrical 1-$\sigma$ error bars on $h$ are computed from the
symmetrical 1-$\sigma$ uncertainties on $h^{1/2}$.

In the discussions below, it is clear that when a detailed examination of
nearby clusters is made, significant departures from the spherically
symmetric, smooth, isothermal cluster ``ideal'' are seen.  Improved X-ray
models from \asca and \rosat are critical to the use of the SZE to determine
$H_0$.

\subsubsection{A478}

Using this X-ray model based upon the Allen \etal \markcite{allen93} (1993) 
\rosat observations and the \ginga temperature, we derive a Hubble
parameter $h^{1/2} = 0.57\pm0.14$ or $h = 0.32{+0.18 \atop -0.14}$.  The
largest contribution to the uncertainty comes from $y_{pred}$ ($\pm24\%$)
rather than the 5.5-m measurement $y_{meas}$ ($\pm7\%$).  
Thus, the largest improvement to be made is in the X-ray model.

The low value of $h = 0.32$ is similar to the value we obtained when
calculating the baryonic masses and mass-fractions in the previous section.
A478 appears to have a much stronger SZE decrement than one would expect from
the X-ray model, as well as a higher X-ray luminosity than one would expect
from the size and velocity dispersion.  One possible explanation for this is
that A478 is significantly elongated along the line of sight, by around a
factor of two compared with its dimensions in the plane of the sky.  This would
bring the implied value of the Hubble constant in line with the other
clusters.  It may also be that A478 is contaminated by the cooling flow
emission.  However, we have done some preliminary tests using a two-component
model incorporating a low-temperature high-density phase, which gives nearly
the same predicted SZE decrement, thus nearly the same Hubble constant.  This
cluster remains a puzzle and merits more detailed examination.

\subsubsection{A2142}

The X-ray model predictions and observed SZE give $h^{1/2} = 0.69\pm0.26$, or
$h = 0.48{+0.43 \atop -0.29}$.  The statistical error is dominated by the the
uncertainty in $y_{pred}$ ($\pm43\%$) rather than in $y_{meas}$ ($\pm6\%$).
The largest single uncertainty is in the value of $\beta$.  It will be
important to improve the model with a detailed analysis of the \rosat data.

\subsubsection{A2256}

Using the Henry \etal\ parameters and the \ginga temperature, combined with
the 5.5-m measurements of the SZE we derive $h^{1/2} = 0.85\pm0.12$, and thus
$h = 0.72{+0.22 \atop -0.19}$.  The contributions to the error bar are
$\pm7\%$ from $y_{pred}$ and $\pm12\%$ from $y_{meas}$.  For this cluster, the
dominant uncertainty is from the SZE measurement.  It will be difficult to
improve these measurements significantly, as it already has over 300 hours of
integration time devoted to it.  In addition, the CMB anisotropies on these
scales are expected to be in the range 14 -- $54\uK$ rms (see
\S~\ref{sze-results}).  Clusters with SZE decrements weaker than that
in A2256 will be very difficult to use for determination of the Hubble
constant.

This cluster appears to have the best model, and cleanest X-ray and SZE data,
though it is weaker than the others.  Some possible problems not apparent in
this analysis may be caused by the presence of the merging sub-clusters in the
core, and the probable presence of very high-temperature gas indicated by
preliminary reports from \asca (J.P. Henry, private communication).  This
should be kept in mind when evaluating the A2256 data (and similarly for the
other clusters), and in the not too distant future \asca and
\rosat should be able to provide much better constraints on the cluster
models.

\subsubsection{Coma}

We will adopt the Herbig \etal \markcite{herbig95} determination of the Hubble
constant, rather than use our own isothermal model (see Table~\ref{tbl:xray}).
Because the Coma cluster subtends a large angle on the sky compared with the
switching angle, there is significant contribution of the SZ decrement to the
reference beams, and thus the details of where the isothermal cluster
atmosphere cuts off makes a noticeable difference to the derived Hubble
constant.

Herbig \etal \markcite{herbig95} find a value of $h^{1/2} = 0.843\pm0.163$,
which after application of the relativistic correction ($\chi_{rel} = 1.029$)
and removal of the $6.9\%$ calibration uncertainty, becomes $h^{1/2} =
0.819\pm0.148$.  This gives us $h = 0.67{+0.26 \atop -0.22}$.  Note that
adoption of a lower temperature, such as the Briel \etal value, will reduce
the derived Hubble constant for Coma.

\subsubsection{Results for the Sample}

If the clusters in our sample are significantly ellipsoidal in shape, a value
for $H_0$ can only be obtained by averaging a number of results from the
unbiased sample.  We combine the measurements for A478, A2142, A2256, and Coma
(using our relativistically corrected Herbig \etal \markcite{herbig95} value),
and the average is listed in Table~\ref{tbl:hubble}.  The most natural
variable to average is $h^{1/2}=y_{pred}/y_{meas}$, for which the measurement
and model errors should enter, as nearly as possible, in a Gaussian fashion,
and for which projection effects (see below) should average out in an
orientation unbiased sample.  In this case, we get the mean
$h^{1/2}=0.733\pm0.076$, or $H_0=54{+12
\atop -11}\kmsm$.  Note that the reduced $\chi^2=0.90$ on the three degrees of
freedom against the mean $h^{1/2}$, so the spread in $H_0$ is consistent with
the (large) error bars.

Up until this point, we have dealt with the ``statistical'' uncertainties
introduced by the observations, calibration, and models.  We should therefore
now include the overall systematic calibration uncertainty of $6.9\%$ (see
\S\ref{obs-cal}).  Because all of the observations were calibrated using 
the same scale (including Coma), any error is correlated between the four
cluster measurements, and should thus be applied to the sample mean as a
whole.  Adding this uncertainty in quadrature, we find a sample average
$h^{1/2}=0.733\pm0.091$, or $H_0=54{+14 \atop -13}\kmsm$.  This is the value
that we adopt.  The reader is reminded also that if the individual cluster
measurements are to be used from Table~\ref{tbl:bmass} or from
Table~\ref{tbl:hubble}, then the $6.9\%$ calibration uncertainty should be
added to the statistical error bars listed there.

Because $H_0$ depends upon the {\it squares} of the $\Delta T$ of the model
and of the measurement, the fractional errors in each are effectively doubled
before adding in quadrature to make up the error budget in the Hubble
constant.  Accurate determination of $H_0$ therefore relies both upon accurate
measurements of the SZE, and upon an accurate model of the state of the
intracluster medium (see discussion in Birkinshaw \etal \markcite{birk91} 
1991, and Inagaki, Suginohara \& Suto \markcite{inagaki95} 1995).

At the beginning of this section, we discussed the systematic errors
introduced by the relativistic corrections to the SZE, and the effect
of a cluster peculiar velocity.  In the former case, corrections to
the $y$-parameter were made, and in the latter case, the corrections 
were dismissed as unlikely to be important.

The most serious potential source of systematic error in the determination of
$y_{pred}$ given the X-ray measurement is elongation of the cluster.  Our
analysis assumes that the line of sight extent of the cluster is the same as
that in the plane of the sky.  We have also assumed a spherical density
profile in our analysis, though we would get the same result for an ellipsoidal
cluster with $r_c$ as the geometric mean core radius.

Deviations from circular symmetry in isophotes are not unusual: McMillan \etal
\markcite{mcmillan88} (1988) studied 49 clusters observed by the {\it
Einstein} satellite and found that the X-ray images had ellipticities of up to
$0.5$.  As discussed in \S~\ref{models-dens}, the X-ray isophotes of A478 and
A2256 show evidence for significant ellipticity.  Elliptical isophotes on the
sky imply, at least statistically, a non-unity axial ratio in the line-of-sight
dimension also.

There is some indication that the cluster A478, with its high implied baryonic
mass fraction and low implied Hubble constant, may be an example of a highly
elongated cluster.  If A478 were excluded from our sample average, then we
would find $h^{1/2}=0.788\pm0.082$, or $H_0=62{+14 \atop -12}\kmsm$.  However,
without any clear indication of a problem with A478 given the large error bars,
we choose to adopt the entire sample average.

It has been found that many clusters have significant cooling cores, and simple
$\beta$-models may be inadequate to describe the state of the gas in these
cases.  Edge and Stewart \markcite{edge91a} (1991a) list A478 and A2142 as
cooling-flow clusters.  Cooling-flow clusters are characterized by cores with
gas at a significantly higher density and lower temperature than the
surrounding gas.  This causes a pronounced central peak in the X-ray surface
brightness of the clusters.  The cooling cores typically have radii of 50--200
kpc and temperatures reduced by up to a factor of four compared to the overall
temperatures.  Because the SZ decrement is more sensitive to the outer, low
density regions of the gas distribution than is the X-ray flux, central
densities calculated from $S_X(r)$ at relatively large radii (Abramopoulos \&
Ku \markcite{abramku83} 1983, Jones \& Forman \markcite{jones84} 1984) should
be used when possible.  These densities are typically 2--3 times lower than the
cooling core densities found by Edge and Stewart \markcite{edge91a} (1991a).
Given X-ray images with high resolution and sensitivity, a better method would
be to model $n_e$ and $T_e$ at large and small radii separately.  Better models
for the gas distribution are needed to account for the presence of cooling
flows or other departures from a single spherically symmmetric smooth 
isothermal profile.

\section{Concluding Remarks} \label{discuss}

In summary, we find significant detections of the SZE for the clusters A478,
A2142 and A2256 using the $5.5$-m telescope at OVRO.  These are the first
observations to detect the effect in these clusters, although A478 and
A2142 have been searched before (Lake \& Partridge \markcite{lake80} 
1980, Birkinshaw \etal \markcite{birk81} 1981, Birkinshaw \& Gull 
\markcite{birk84} 1984, Chase \etal \markcite{chase87} 1987).  Observations of
contaminating radio sources were carried out on the OVRO 40-m telescope.  
When the SZE measurements of the X-ray flux-limited sample are complete,
an orientation unbiased sample of clusters will be available for measurement
of the Hubble constant.

The SZE is a measure of the electron pressure in the ionized cluster medium
--- with knowledge of the electron temperature, the SZE is proportional to the
baryonic mass in the IGM contained within the telescope beam.  We find similar
baryonic mass surface densities for the three clusters and Coma: $\Sigma_g
\sim 7 \times 10^{13} M_\odot\,{\rm Mpc}^{-2}$.  For A2142, A2256 and Coma,
consistent estimates of the baryonic mass fraction $M_{sze}/M_{tot} \approx
0.061\pm0.010\,h^{-1}$.  This is a lower limit on the total baryonic mass,
as the galaxies contribute $M_{sze}/M_{tot} \approx 0.009 \pm 0.003$.
When compared with the standard primordial
nucleosynthesis estimates for $\Omega_B$, we find consistency between the SZE
data and nucleosynthesis for cosmological density parameters in the range
$\Omega_0 h \approx 0.21\pm0.05$.  This agrees with the values determined
independently from large-scale structure and galaxy counts.  The cluster A478
gives a discrepant (high) fraction of $M_{sze}/M_{tot} \sim 0.17\,h^{-1}$, and
is likely elongated or heavily contaminated by the cooling flow (or both).

Recent determinations of $\Omega_B$ using the deuterium abundances in
Lyman-$\alpha$ absorption systems along the line-of-sight to QSOs give
discrepant values higher and lower than the standard.  If we adopt a high
baryon density $\Omega_B h^2 = 0.024\pm0.006$ (Tytler, Fan and Burles
\markcite{tytler96} 1996), then our data imply $\Omega_0 h \leq 0.39\pm0.12$.
On the other hand, if $\Omega_B h^2 = 0.0062 \pm 0.0008$ (Rugers \& Hogan
\markcite{hogan96} 1996), then $\Omega_0 h \leq 0.10\pm0.02$.

By combining the measured SZE decrements with published X-ray models, we have
determined the value of the Hubble constant implied for each of these
clusters.  Clusters A478 and A2142 are complicated by excess core emission
attributed to cooling flows, and better X-ray models for the gas distribution
must be obtained.  A2256 appears to be undergoing a merging event, and
additional modeling must also be done here.  With the preliminary models
gleaned from the literature, we obtain an average 
of $H_0=54{+14 \atop -13}\kmsm$ for A478, A2142, A2256, and Coma.
This average value tends toward the low side of the commonly accepted range,
as do most of the other SZE determined values (\eg Birkinshaw \etal 
\markcite{birk91} 1991, Jones \etal \markcite{jones93} 1993, Birkinshaw
\& Hughes \markcite{birk94} 1994), though the large error bars
place our measurement within $3\,\sigma$ of practically all of the other 
values for $H_0$.  

Note that the high baryon density of Tytler, Fan and Burles
\markcite{tytler96} (1996), our measurement of the baryon fraction, and our
average value for the Hubble constant would imply a high overall density for
the Universe $\Omega_0 \leq 0.72\pm0.29$.  This is marginally consistent with
a Universe with the critical density $\Omega_0 = 1$.  However, adoption of the
lower values of the baryon density would favor the acceptance of a low density
parameter.

Possible problems with using the SZE and X-ray measurements for inferring
$H_0$ include substructure in cluster atmospheres, elongation of clusters
along the line of sight, and the presence of cooling cores.  The SZE decrement
is sensitive to the outer regions of the intracluster gas, which have not been
well studied because of their relatively faint X-ray emission.  Better models
of the cluster atmospheres will soon be provided by the new generation of
X-ray satellites, such as \asca and AXAF.

The advantage of determining $H_0$ from a well-selected sample of clusters is
the ability to use the distribution of derived $H_0$ to test for variations in
the astrophysical parameters of the cluster models assumed in the analysis.
As this stage, our models are too uncertain and therefore our error bars too
large, to assess any but the grossest deviations in derived $h^{1/2}$.  The
most discrepant value is $H_0=32{+18 \atop -14}\kmsm$ for A478, and even this
is less than $2\sigma$ from the mean.  If
massive clusters are inordinately elongated, estimates of $H_0$ from
individual clusters may be off by a factor of two or more.  Exclusion of A478
from the sample average raises the value of $H_0$ somewhat, though this step
is unwarranted by the data at hand.  

The ability to recognize deviant clusters like A478 demonstrates the power of
using this sample of clusters.  Clearly, completing the entire sample is the
proper way to proceed using this method.  However, the clusters reported
here were selected as the first targets because they were free from strong
contamination by radio sources.  It will be very difficult to measure the
SZE in the remaining clusters with the accuracy that we have been able to
achieve with these first results.  

In the end, it will be the distribution of the $H_0$ values for the sample
that will tell us whether clusters are suitable tools with which to measure
the expansion of Universe, or whether variations in shape and orientation, or
density and temperature substructure introduce severe limitations in the
determination of $H_0$ by this method.  Individual clusters are insufficient
to make the case for one value of $H_0$ or another, and it remains to be
demonstrated that this method will yield reliable results.  In any event,
sounding of the intragalactic medium through combined SZE and X-ray
measurements promises to provide important constraints upon multi-phase
structures in the hot cluster atmosphere.  This is as important as determining
$H_0$, in that a number of cosmological tests, such as the baryon fraction
$\Omega_B$, rely upon observations of clusters of galaxies.

\acknowledgments

STM was supported by a R.A.\ Millikan Fellowship at Caltech.  Experimental
Microwave Astrophysics at Caltech is supported by NSF grant AST-9117100 and
AST-9419279.  STM also acknowledges the hospitality the ITP in Santa Barbara,
which is supported by the NSF under grant PHY89-04035.  The observations
at the Owens Valley Radio Observatory could not have been made without the
expert help of Harry Hardebeck, Mark Hodges, and Russ Keeney.  We also thank
Tim Pearson for creating the control software for the OVRO 5.5-m and 40-m
telescopes and for many useful discussions.

\def\phb{\phantom{00}}

\clearpage

\begin{deluxetable}{l l l c c c c c }
\tablecaption{ The X-ray Selected Sample \label{tbl:sample} }
\tablewidth{0pt}
\tablecolumns{8}
\tablehead{ \sc Cluster 
 & \multicolumn{2}{c}{Position (B1950)} & \colhead{$z$} 
 & \colhead{$f_x/10^{-11}$} & \colhead{$L_x/10^{44}$}
 & \colhead{$kT$} & \colhead{$\theta_{core}$} \\ 
\rm  & \colhead{R.A.} & \colhead{Dec.} & \colhead{ }
 & \colhead{ $\rm erg\, cm^{-2}\, s^{-1}$ }
 & \colhead{ $h^{-2}\,{\rm erg\, s^{-1}}$ }
 & \colhead{(keV)} & \colhead{(arcmin)} \\
 \colhead{(Notes)} & \colhead{[1]} & \colhead{[1]} & \colhead{[2]} 
 & \colhead{[2]}   & \colhead{[2]} & \colhead{[3]} & \colhead{[4]} }
\startdata
A85   & 00:39:19.5 &$-$09:34:23 & 0.0518 & \phn6.37 & 1.88 & 6.2 & 
 \phn2\farcm51 \nl
A399  & 02:55:07.6 &  +12:50:47 & 0.0715 & \phn3.41 & 1.94 & 5.8 &
 \phn1\farcm91 \nl
A401  & 02:56:12.0 &  +13:22:43 & 0.0748 & \phn5.88 & 3.68 & 7.8 &
 \phn4\farcm68 \nl
A478  & 04:10:40.1 &  +10:20:21 & 0.0900 & \phn6.63 & 6.02 & 6.6 &
 \phn1\farcm85 \nl
A754  & 09:06:49.7 &$-$09:28:57 & 0.0528 & \phn8.53 & 2.62 & 9.1 &
 \phn8\farcm45 \nl
A1651 & 12:56:48   &$-$03:55:00 & 0.0825 & \phn3.67 & 2.80 & 7.0 &
 \nodata \nl
Coma  & 12:57:19   &  +28:13:24 & 0.0232 & 25.4\phn & 1.49 & 8.1 &
 10\farcm5\phn \nl
A1795 & 13:46:35.4 &  +26:50:23 & 0.0616 & \phn5.30 & 2.23 & 5.3 &
 \phn3\farcm03 \nl
A2029 & 15:08:27.2 &  +05:55:56 & 0.0767 & \phn7.52 & 4.92 & 7.8 &
 \phn1\farcm58 \nl
A2142 & 15:56:15.8 &  +27:22:38 & 0.0899 & \phn7.50 & 6.80 & 8.7 &
 \phn3\farcm69 \nl
A2256 & 17:06:56.3 &  +78:43:02 & 0.0601 & \phn5.20 & 2.08 & 7.5 &
 \phn5\farcm33 \nl
\enddata

\tablecomments{ [1] Positions from {\it Einstein } IPC, except A1651 from 
Abell, \etal \markcite{abell89} (1989).  [2] Redshifts, X-ray fluxes and
luminosities (2-10 keV) from Edge \etal \markcite{edge90} (1990).
Luminosities assume $q_0=1/2$.  [3] X-ray temperatures from
\eins MPC (David \etal \markcite{david93} 1993) except: A478 from \ginga and
\rosat (Allen \etal \markcite{allen93} 1993); A1651 from \heao1 (Edge \etal
\markcite{edge90} 1990); Coma, A1795, A2142, A2256 from \ginga (David \etal 
\markcite{david93} 1993).  [4] Core radii from Jones \& Forman 
\markcite{jones84} (1984) except: A401, A754, A2142 from Abramopoulos \& 
Ku (1983); A478 from Allen \etal \markcite{allen93} (1993); Coma from 
Briel \etal \markcite{briel92} (1992); A2256 from Henry \etal 
\markcite{henry93} (1993). }

\end{deluxetable}

\clearpage

\begin{deluxetable}{ l l l }
\tablecaption{ Pointing Positions for SZE Observations \label{tbl:pos} }
\tablewidth{0pt}
\tablecolumns{3}
\tablehead{ \sc Cluster & \multicolumn{2}{c}{Field Position (J2000)} \\ 
\rm & \colhead{R.A.} & \colhead{Dec.} }
\startdata
A478L  & 03:54:25.01 & +10:27:40.73 \nl 
A478   & 04:13:25.01 & +10:27:40.73 \nl
A478T  & 04:32:25.01 & +10:27:40.73 \nl \tablevspace{1ex} 
A2142L & 15:43:18.00 & +27:13:32.00 \nl 
A2142  & 15:58:18.00 & +27:13:32.00 \nl 
A2142T & 16:13:18.00 & +27:13:32.00 \nl \tablevspace{1ex} 
A2256L & 16:48:54.50 & +78:38:27.00 \nl 
A2256  & 17:03:54.50 & +78:38:27.00 \nl 
A2256T & 17:18:54.50 & +78:38:27.00 \nl
\enddata
\end{deluxetable}

\begin{deluxetable}{l r@{$\pm$}l r@{$\pm$}l r@{$\pm$}l r@{$\pm$}l r@{$\pm$}l}
\tablecaption{ OVRO 5.5-m Standard Calibration Scale \label{tbl:cals} }
\tablewidth{0pt}
\tablehead{ \colhead{\sc Scale}
   & \multicolumn{2}{c}{$T_{Mars}$ } 
   & \multicolumn{2}{c}{$S_{DR21}$ }
   & \multicolumn{2}{c}{$S_{N7027}$ } 
   & \multicolumn{2}{c}{$S_{3C286}$ } }
\startdata
Relative to Jupiter  & 1.252&0.022 & 134.7&2.1 mJy/K & 35.5&1.1 mJy/K 
 & 13.2&0.6 mJy/K \nl
$T_{Jupiter} = 144\pm8$ K  &   180&10 K  &  19.4&1.1 Jy    & 5.11&0.33 Jy   
 & 1.90&0.14 Jy \nl
\enddata
\end{deluxetable}

\begin{deluxetable}{l c c c}
\tablecaption{ OVRO 5.5-m Measurements of SZE \label{tbl:sze} }
\tablewidth{0pt}
\tablecolumns{4}
\tablehead{  & \colhead{A478} & \colhead{A2142} &  \colhead{A2256} \\
\rm & \colhead{($\uK$)} & \colhead{$\uK$)} & \colhead{($\uK$)} }
\startdata
M$-$(L+T)/2  & $-375\pm24$    & $-420\pm19$    & $-218\pm14$    \nl
M$-$L        & $-366\pm27$    & $-399\pm20$    & $-217\pm15$    \nl
M$-$T        & $-386\pm27$    & $-451\pm22$    & $-238\pm16$    \nl
\tablevspace{1ex}
MAIN         & $-139\pm20$    & $-214\pm17$    & $-310\pm12$    \nl
LEAD         & \phs$244\pm21$ & \phs$159\pm17$ & \phn$-94\pm12$ \nl
TRAIL        & \phs$242\pm20$ & \phs$226\pm16$ & \phn$-82\pm12$ \nl
L$-$T        & \phn$-13\pm27$ & \phn$-66\pm20$ & \phn$-22\pm15$ \nl
(L$+$T)/2    & \phs$248\pm15$ & \phs$188\pm13$ & \phn$-74\pm13$ \nl
\tablevspace{1ex}
$\nu_{eff}$  & \phn800        & \phn840        &    2020 \nl  
n$_{pts}$    &    1146        &    1337        &    4117 \nl  
$\tau$ (hrs) &  \phb64        &  \phb85        & \phn310 \nl  
\enddata
\end{deluxetable}

\clearpage

\begin{deluxetable}{l l l r@{$\pm$}l c c c c}
\tablecaption{ Sources within $9\arcm$ of field centers or
   reference arcs. \label{tbl:src40m} } 
\tablewidth{0pt}
\tablecolumns{7}
\tablehead{\sc Source & \multicolumn{2}{c}{Position (J2000)} 
  & \multicolumn{2}{c}{$S_{low}$} & \colhead{$S_{18.5}$} & \colhead{$\alpha$} 
  & \colhead{$S_{32}$} \\
\rm & \colhead{R.A.} & \colhead{Dec.} 
  & \multicolumn{2}{c}{(mJy) \tablenotemark{a}} & 
  \colhead{(mJy)} & \colhead{ } & \colhead{(mJy)} }
\startdata
A478L.1  & 03:54:57.2 & $+$10:12:30 & 33&7 & $<3.9$ & $<-1.4$ & $<1.8$ \nl
A478.1   & 04:13:34   & $+$10:28:04 & 16&7 \tablenotemark{b} & $<9.1$ 
  & $< 0.2$ & $<9.2$ \nl 
\tablevspace{1ex} 
A2142L.1 & 15:41:46.8 & $+$27:05:54 & 51&8 & 13.5$\pm$1.1 & $-0.99\pm0.13$ 
  & 7.8$\pm$0.9 \nl
A2142L.2 & 15:42:58.4 & $+$27:06:46 & 34&7 & $<5.1$ & $<-1.2$ & $<2.6$ \nl 
A2142.1  & 15:57:11.2 & $+$26:51:31 & 52&9 & \phn8.5$\pm$1.9 & $-1.35\pm0.21$ 
  & 4.1$\pm$1.1 \nl 
A2142.2  & 15:58:14.3 & $+$27:15:48 & 44&8 & \phn7.9$\pm$2.0 & $-1.28\pm0.23$ 
  & 3.9$\pm$1.1 \nl 
A2142.3  & 15:58:47   & $+$27:18:06 & 18&7 \tablenotemark{c} & $<6.3$ 
  & $<-0.3$ & $<5.4$ \nl
A2142.4  & 15:59:05   & $+$27:03:19 & 42&15 \tablenotemark{c} & $<5.1$
  & $<-0.9$ & $<3.2$ \nl
A2142T.1 & 16:12:26.3 & $+$27:23:16 & 49&8 & 14.3$\pm$1.5 & $-0.92\pm0.14$ 
  & 8.6$\pm$1.1 \nl  
\tablevspace{1ex} 
A2256.1  & 17:02:09   & $+$78:40:56 &  48&8  \tablenotemark{d} & $<6.9$ 
  & $<-0.7$ & $<4.8$ \nl 
A2256.2  & 17:02:28   & $+$78:42:57 & 166&11 \tablenotemark{d} & $<4.8$ 
  & $<-1.3$ & $<2.3$ \nl
A2256.3  & 17:03:03   & $+$78:36:40 &  62&4  \tablenotemark{d} & $<5.1$ 
  & $<-0.9$ & $<3.0$ \nl 
A2256.4  & 17:03:09   & $+$78:40:00 &  39&3  \tablenotemark{d} & $<6.0$
  & $<-0.7$ & $<4.1$ \nl
A2256.5  & 17:03:28   & $+$78:37:58 & 157&10 \tablenotemark{d} 
  & \phn8.3$\pm$2.5 & $-1.14\pm0.12$ & 4.4$\pm$1.4 \nl
A2256.6  & 17:03:51   & $+$78:46:03 & 185&13 \tablenotemark{d} & $<10.8$
  & $<-1.1$ & $<6.0$ \nl
A2256.7  & 17:04:48.9 & $+$78:38:29 &  11&1  \tablenotemark{d} & $<10.5$ 
  & $<0.03$ & $<10.7$ \nl 
\enddata
\tablenotetext{a}{Flux density at $4.85$ GHz from 87GB unless otherwise noted.}
\tablenotetext{b}{Flux density at $10.7$ GHz (Andernach \etal 1986). }
\tablenotetext{c}{Flux density at $2.7$ GHz (Andernach \etal 1986). }
\tablenotetext{d}{Flux density at $1.415$ GHz (Bridle \etal 1979). } 
\end{deluxetable}

\clearpage 

\begin{deluxetable}{l c c c c c c }
\tablecaption{ Corrections for sources within $9\arcm$ of field centers or
   reference arcs. \label{tbl:src1} }
\tablewidth{0pt}
\tablecolumns{7}
\tablehead{ \sc Source & \colhead{\sc Rad} & \colhead{\sc Pos} 
  & \colhead{$d$} & \colhead{$\psi_p$} & \colhead{$\Delta T_a$} 
  & \colhead{$\Delta T_{max}$} \\
\rm  & \colhead{ } & \colhead{\sc Ang} & \colhead{ } & \colhead{ } & 
    \colhead{$(\uK)$} & \colhead{$(\uK)$} }
\startdata
A478L.1  & $17\farcm12$ & $152^\circ$ & $5\farcm81$ & \phs$62^\circ$ 
  & $<11$ & $<1$ \nl
A478.1   & \phn$2\farcm24$ & \phn$80^\circ$ & \nodata & \nodata & $<57$ 
  & $<44$ \nl
\tablevspace{1ex} 
A2142L.1 & $21\farcm67$ & $249^\circ$ & $0\farcm49$ & $-21^\circ$ 
  & 48$\pm$6 & $-24\pm3$ \nl
A2142L.2 & \phn$8\farcm05$ & $213^\circ$ & \nodata & \nodata & $<16$ & $<1$ \nl
A2142.1  & $26\farcm57$ & $214^\circ$ & $4\farcm41$ & $-56^\circ$ 
  & 26$\pm$7 & \phn$-5\pm1$ \nl
A2142.2  & \phn$2\farcm41$ & $340^\circ$ & \nodata & \nodata & 24$\pm$7 
  & \phs$18\pm5$ \nl
A2142.3  & \phn$7\farcm90$ & \phn$55^\circ$ & \nodata & \nodata & $<34$ 
  & $<1$ \nl
A2142.4  & $14\farcm62$ & $134^\circ$ & $7\farcm54$ & \phs$44^\circ$ & $<20$ 
  & $<1$ \nl
A2142T.1 & $15\farcm05$ & $310^\circ$ & $7\farcm11$ & \phs$40^\circ$ & 53$\pm$7
  & $-2.0\pm0.3$ \nl  
\tablevspace{1ex} 
A2256.1  &\phn$5\farcm75$ & $296^\circ$ & \nodata & \nodata & $<30$ & $<5$ \nl
A2256.2  &\phn$6\farcm19$ & $317^\circ$ & \nodata & \nodata & $<14$ & $<2$ \nl
A2256.3  &\phn$3\farcm10$ & $235^\circ$ & \nodata & \nodata & $<19$ & $<11$ \nl
A2256.4  &\phn$2\farcm72$ & $305^\circ$ & \nodata & \nodata & $<26$ & $<17$ \nl
A2256.5  &\phn$1\farcm39$ & $250^\circ$ & \nodata & \nodata & 27$\pm$9 
  & \phs$25\pm8$ \nl
A2256.6  &\phn$7\farcm60$ & $359^\circ$ & \nodata & \nodata & $<37$ & $<2$ \nl
A2256.7  &\phn$2\farcm68$ & \phn$89^\circ$ & \nodata & \nodata & $<67$ 
  & $<46$ \nl
\enddata
\tablecomments{ Radius (arcmin) and Position Angle are measured from center of 
main field. Distance $d$ (arcmin) is measured from center of reference arc at 
parallactic angle $\psi_p$. }
\end{deluxetable}

\begin{deluxetable}{l c c c c c }
\tablecaption{ Total source contributions to $5.5$-m SZE Fields
\label{tbl:src5m} }
\tablewidth{0pt}
\tablecolumns{6}
\tablehead{ \sc Field & \colhead{$N_{on}$} & \colhead{$\Delta T_{on}$} 
  & \colhead{$N_{ref}$} & \colhead{ max $\Delta T_{ref}$ } 
  & \colhead{$\Delta T_{fld}$} \\
\rm & \colhead{ } & \colhead{($\uK$)} & \colhead{ } & \colhead{($\uK$)} 
  & \colhead{($\uK$)} }
\startdata
A2142L & 1 & \phn0.6  & 1 &     $-$23.7 & \phn$-$0.7 \nl
A2142  & 1 &    17.8  & 1 &  \phn$-$4.8 & \phs16.2   \nl
A2142T & 0 & \phn0.0  & 1 &  \phn$-$2.0 & \phn$-$0.2 \nl
\tablevspace{1ex} 
A2256  & 1 &    24.5  & 0 & \phs\phn0.0 & \phs24.5   \nl
\enddata 
\end{deluxetable}

\begin{deluxetable}{l c c l }
\tablecaption{ Final SZE Results \label{tbl:results} }
\tablewidth{0pt}
\tablecolumns{4}
\tablehead{ \sc Cluster  
   & \colhead{$\Delta T_{\rm 5.5m}(\mbox{obs})$} 
   & \colhead{$\Delta T_{\rm 5.5m}(\mbox{src})$} 
   & \colhead{$\Delta T_{\rm 5.5m}(\mbox{corr})$} 
   \\
\rm & \colhead{($\uK$)} 
   & \colhead{($\uK$)}
   & \colhead{($\uK$)} }
\startdata
A478  & $-375\pm24$ & $<$44 & $-375\pm28$ \tablenotemark{a} \nl
A2142 & $-420\pm19$ & \phantom{$<$}17 & $-437\pm25$ \nl
A2256 & $-218\pm14$ & \phantom{$<$}25 & $-243\pm29$ \nl
\enddata
\tablenotetext{a}{ A478 source limit $3\sigma$, used $1\sigma$ for
 uncertainty.}
\end{deluxetable}

\begin{deluxetable}{l c c c c c c c}
\tablecaption{ X-Ray Cluster Parameters \label{tbl:xray} }
\tablewidth{0pt}
\tablecolumns{7}
\tablehead{ \sc Cluster & $z$ & \colhead{$\theta_{core}$} & \colhead{$kT_e$} 
 & \colhead{$n_0/10^{-3}$}   & \colhead{$\beta$} & \colhead{$r_c$} 
 \\ 
\rm & \colhead{ } & \colhead{(arcmin)} & \colhead{(keV) } 
 & \colhead{($h^{1/2}\,{\rm cm^{-3}}$)} & \colhead{ } 
 & \colhead{($h^{-1}$ Mpc) \tablenotemark{a}} }
\startdata
A478 & 0.0881 & \phn1.93$\pm$0.30 & 6.56$\pm$0.09 &  9.55$\pm$1.75 
  & 0.667$\pm$0.029 & 0.128$\pm$0.020 \nl 
A2142& 0.0899 & \phn3.69$\pm$0.14 & 8.68$\pm$0.12 &  6.97$\pm$0.41 
  & \phb1.0$\pm$0.3\phb & 0.249$\pm$0.009 \nl 
A2256& 0.0581 & \phn5.33$\pm$0.20 & 7.51$\pm$0.11 &  3.55$\pm$0.18 
  & 0.795$\pm$0.020 & 0.245$\pm$0.009 \nl
Coma & 0.0235 &    10.50$\pm$0.60 & 9.10$\pm$0.40 &  4.09$\pm$0.06 
  & 0.750$\pm$0.030 & 0.207$\pm$0.012 \nl 
\enddata
\tablenotetext{a}{Assumes $q_0=1/2$.}
\tablecomments{ X-ray temperatures from \ginga (David \etal \markcite{david93}
 1993), except Coma (Hughes, Gorenstein \& Fabricant \markcite{hughes88}
 1988). Other parameters: A478 from Allen \etal \markcite{allen93}(1993) and
our own analysis of \rosat data; A2142 from Abramopoulos \& Ku
\markcite{abramku83} (1983); A2256 from Henry \etal \markcite{henry93} (1993);
Coma $r_c$, $n_0$ and $\beta$ from Briel, Henry \& B\"{o}hringer
\markcite{briel92}(1992). }
\end{deluxetable}

\begin{deluxetable}{l c c c c }
\tablecaption{ Efficiencies and Relativistic Corrections \label{tbl:relcor} }
\tablewidth{0pt}
\tablecolumns{5}
\tablehead{ \sc Cluster & \colhead{$\eta_g$} & \colhead{$\eta_{obs}$} &
   \colhead{$\chi_{rel}$} & \colhead{$y_{meas}$} \\
\rm & \colhead{ } & \colhead{ } & \colhead{ } & \colhead{($10^{-5}$)} }
\startdata
 A478   & 0.826$\pm$0.028 & 0.414$\pm$0.008 & 1.022 & 7.41$\pm$0.55 \nl
 A2142  & 0.947$\pm$0.081 & 0.498$\pm$0.090 & 1.030 & 8.70$\pm$0.50 \nl
 A2256  & 0.818$\pm$0.014 & 0.612$\pm$0.003 & 1.026 & 4.82$\pm$0.58 \nl
 Coma   & 0.618$\pm$0.026 & 0.563$\pm$0.021 & 1.029 & 6.13$\pm$0.93 \nl
\enddata
\end{deluxetable}

\begin{deluxetable}{l c c c }
\tablecaption{ Baryonic Masses from the SZE \label{tbl:bmass} }
\tablewidth{0pt}
\tablecolumns{4}
\tablehead{ \sc Cluster & \colhead{$L_g$} & \colhead{$\Sigma_g$} 
   & \colhead{$M_g$} \\
\rm & \colhead{($h^{-1}$ Mpc) \tablenotemark{a} }
   & \colhead{($10^{13}\,M_\odot\,{\rm Mpc}^{-2}$)}  
   & \colhead{($10^{13}\,h^{-2}\,M_\odot$)} }
\startdata
 A478   & 0.207\phn & $9.59\pm0.79$ & $2.58\pm0.21$ \nl
 A2142  & 0.211\phn & $7.42\pm0.77$ & $2.08\pm0.22$ \nl
 A2256  & 0.143\phn & $5.50\pm0.67$ & $0.71\pm0.09$ \nl
 Coma   & 0.0615    & $7.64\pm1.25$ & $0.18\pm0.03$ \nl
\enddata
\tablenotetext{a}{Assumes $q_0=1/2$.}
\end{deluxetable}

\begin{deluxetable}{l c c c c c }
\tablecaption{ Baryonic Fractions in the Clusters \label{tbl:bfrac} }
\tablewidth{0pt}
\tablecolumns{7}
\tablehead{ \sc Cluster & \colhead{$R_0$} & \colhead{$V_s(R_0)/V_g$}
   & \colhead{$M_{sze}$} & \colhead{$M_{tot}$} & \colhead{$M_{sze}/M_{tot}$}
   \\
\rm 
   & \colhead{($h^{-1}$ Mpc)} & \colhead{ } 
   & \colhead{($10^{13}\,h^{-2}\,M_\odot$)}
   & \colhead{($10^{13}\,h^{-1}\,M_\odot$)}
   & \colhead{($h^{-1}$)} }
\startdata
 A478 \tablenotemark{a} & 0.976 & 2.99$\pm$0.08 & $7.7\pm0.7$ 
  & \phn46.4 & $0.166\pm0.014$ \nl
 A2142 \tablenotemark{a}& 0.966 & 2.90$\pm$0.43 & $6.0\pm1.1$ 
  & 100.5 & $0.060\pm0.011$ \nl
 A2256 \tablenotemark{b}& 0.76\phn & 4.29$\pm$0.05 & $3.0\pm0.4$ 
  &\phn$51\pm14$ & $0.060\pm0.018$ \nl
 Coma \tablenotemark{c} & 1.50\phn & 38.3$\pm$3.0\phn & $6.9\pm1.3$ 
  & $110\pm22$ & $0.063\pm0.017$ \nl
\enddata
\tablenotetext{a}{A478 and A2142 $M_{tot}$ from White \& Fabian 
  \markcite{white95} (1995). No uncertainties given.}
\tablenotetext{b}{A2256 $M_{tot}$ from Henry, Briel \& Nulsen 
  \markcite{henry93} (1993).}
\tablenotetext{c}{Coma $M_{tot}$ from White \etal 
  \markcite{white93} (1993).}
\end{deluxetable}

\begin{deluxetable}{l c c c c}
\tablecaption{ Hubble Constant from the SZE \label{tbl:hubble} }
\tablewidth{0pt}
\tablecolumns{5}
\tablehead{ \sc Cluster & \colhead{$y_{meas}$} & \colhead{$y_{pred}$} 
   & \colhead{$h^{1/2}$} & \colhead{$H_0$} \\
\rm & \colhead{($10^{-5}$)} & \colhead{($10^{-5}\,h^{-1/2}$)} 
   & \colhead{ } & \colhead{(km s$^{-1}$ Mpc$^{-1}$)} }
\startdata
 A478   & $7.4\pm0.6$ & $4.2\pm1.0$ & $0.57\pm0.14$ & $32{+18 \atop -14}$ \nl
 A2142  & $8.7\pm0.5$ & $6.0\pm2.2$ & $0.69\pm0.26$ & $48{+43 \atop -29}$ \nl
 A2256  & $4.8\pm0.6$ & $4.1\pm0.3$ & $0.85\pm0.12$ & $72{+22 \atop -19}$ \nl
 Coma   & $6.1\pm0.9$ &  \nodata    & $0.82\pm0.15$ & $67{+26 \atop -22}$ \nl
\tablevspace{2ex} 
\sc Sample & \nodata  &  \nodata    & $0.73\pm0.08$ & $54{+12 \atop -11}$ \nl
\enddata
\end{deluxetable}

\clearpage


\begin{figure}[t]
\caption{\label{fig:edit} 
\sf The effect of the data editing and filtering method upon the final
referenced results.  We show results for twenty different editing methods, for
both unweighted (dashed lines) and weighted (solid lines) averages.  The
points are plotted slightly displaced from the filter method.  The methods are
roughly in order of fraction of data accepted, ranging from 100\% (method 1)
to 44\% (method 20).  The effective fraction of data used in the weighted
averages ranges from 37\% to 27\% across the filter methods.  We adopt method
15 with weighting (marked with the triangle), in which $57\%$ of the data is
retained, with an effective weighted fraction of 31\%.  }
\end{figure}

\begin{figure}[t]
\caption{\label{fig:a478} 
\sf Dependence of the LEAD, MAIN, and TRAIL (upper) and MLT and L-T referenced
data (lower) on parallactic angle are shown for A478.  In the upper panels, the
LEAD fluxes are offset to the left of the proper $\psi_p$ and the TRAIL fluxes
are offset to the right.  In the lower panel, the solid and dotted horizontal
lines depict the means of the MLT and L-T.  No source contributions were 
subtracted from A478.}
\end{figure}

\begin{figure}[t]
\caption{\label{fig:a2142} 
\sf Dependence of the LEAD, MAIN, and TRAIL (upper) and MLT and L-T referenced
data (lower) on parallactic angle are shown for A2142.  In the lower panel,
the solid and dotted horizontal curves represent response of the MLT and L-T
data to the contaminating sources.  Note the signal at $\psi_p=-21^\circ$ that
matches the deviation in the data.  The expected signal from the source is
below that observed, and we may have underestimated its contribution.
However, because the number of data points in this $\psi_p$ range is small,
this does not affect the results significantly.}
\end{figure}

\begin{figure}[t]
\caption{\label{fig:a2256} \sf Dependence of the LEAD, MAIN, and TRAIL (upper)
and MLT and L-T referenced data (lower) on parallactic angle are shown for
A2256.  In the lower panel, the dotted horizontal line depicts the expected
zero mean L-T, as no scans of these fields were available.  The dotted 
line is the MLT mean before main beam source correction, and the solid 
line is the mean after source correction.  }
\end{figure}

\clearpage

\begin{figure}[t]
\figurenum{1}
\epsfbox[70 144 630 718]{filter.eps}
\caption{The dependence of SZE results on data editing.} 
\end{figure}

\begin{figure}[t]
\figurenum{2}
\epsfbox[70 144 630 718]{a478rev.eps}
\caption{SZE in A478.}
\end{figure}

\begin{figure}[t]
\figurenum{3}
\epsfbox[70 144 630 718]{a2142rev.eps}
\caption{SZE in A2142.} 
\end{figure}

\begin{figure}[t]
\figurenum{4}
\epsfbox[70 144 630 718]{a2256rev.eps}
\caption{SZE in A2256.} 
\end{figure}

\end{document}